\documentclass[twocolumn]{aastex6}

\pdfoutput=1

\newcommand{\Kepler}{\emph{Kepler}}

\newcommand{\mathbfit}[1]{\textbf{\textit{#1}}}

\newcommand{\mathbfss}[1]{\textbf{\textsf{#1}}}

\makeatletter
\newsavebox\myboxA
\newsavebox\myboxB
\newlength\mylenA

\newcommand*\xoverline[2][0.75]{%
    \sbox{\myboxA}{$\m@th#2$}%
    \setbox\myboxB\null
    \ht\myboxB=\ht\myboxA%
    \dp\myboxB=\dp\myboxA%
    \wd\myboxB=#1\wd\myboxA
    \sbox\myboxB{$\m@th\overline{\copy\myboxB}$}
    \setlength\mylenA{\the\wd\myboxA}
    \addtolength\mylenA{-\the\wd\myboxB}%
    \ifdim\wd\myboxB<\wd\myboxA%
       \rlap{\hskip 0.5\mylenA\usebox\myboxB}{\usebox\myboxA}%
    \else
        \hskip -0.5\mylenA\rlap{\usebox\myboxA}{\hskip 0.5\mylenA\usebox\myboxB}%
    \fi}
\makeatother

\usepackage{amsmath}
\usepackage{amsfonts}
\usepackage{amssymb}
\usepackage{wasysym}
\usepackage[percent]{overpic}
\usepackage{graphicx}

\makeatletter
\def\env@matrix{\hskip -\arraycolsep 
  \let\@ifnextchar\new@ifnextchar
  \array{*{\c@MaxMatrixCols}c}}
\makeatother

\begin{document}

\title{Shadow Imaging of Transiting Objects}

\author{Emily Sandford\altaffilmark{1} and David Kipping}
\affil{Department of Astronomy\\
Columbia University \\
550 W 120th St. \\
New York, NY 10027, USA}

\altaffiltext{1}{esandford@astro.columbia.edu}

\begin{abstract}

We consider the problem of inferring the shape of a transiting object's silhouette from its light curve alone, without assuming a physical model for the object. We model the object as a grid of pixels which transits the star; each pixel has an opacity, ranging from transparent to opaque, which we infer from the light curve. We explore three interesting degeneracies inherent to this problem, in which markedly different transiting shapes can produce identical light curves: (i) the ``flip" degeneracy, by which two pixels transiting at the same impact parameter on opposite sides of the star's horizontal midplane generate the same light curve; (ii) the ``arc" degeneracy, by which opacity can be redistributed along the semicircular arc of pixels which undergoes ingress or egress at the same time without consequence to the light curve, and (iii) the ``stretch" degeneracy, by which a wide shape moving fast can produce the same light curve as a narrow shape moving more slowly. By understanding these degeneracies and adopting some additional assumptions, we are able to numerically recover informative shadow images of transiting objects, and we explore a number of different algorithmic approaches to this problem. We apply our methods to real data, including the TRAPPIST-1c,e,f triple transit and two dips of Boyajian's Star. We provide \texttt{Python} code to calculate the transit light curve of any grid and, conversely, infer the image grid which generates any light curve in a software package accompanying this paper, \href{https://github.com/esandford/EightBitTransit}{\texttt{EightBitTransit}}.

\end{abstract}


\section{INTRODUCTION}
\label{sec:intro}

Transit light curves are rich in information. If we assume a physical model for a transiting object---usually, a spherical body in a Keplerian orbit about a host star---we may then infer the parameters of this model, including physical properties of the transiter, its orbit, and the host star, from the light curve. 

However, anomalous transit-like events, such as those observed in star KIC 8462852 \citep{boyajian16}, resist this type of analysis, because their physical cause, and consequently the appropriate model, is not apparent. In this paper, we consider the general problem of inferring the transiting shape, or shadow image, that generated a particular light curve. We wish to infer this image from the light curve alone, with as few additional assumptions as possible.

A number of problems related to shadow imaging have been studied before. The inverse problem, of how to calculate the light curve of an arbitrary transiting shape, has been tangentially addressed by several numerical transit-light-curve-calculating codes. Generally, however, these assume some parametric model for the transiting object---in the case of \texttt{BATMAN} \citep{kreidberg15}, the transiting object must be a spherical planet; \texttt{LUNA} \citep{kipping11}, a spherical planet accompanied by a spherical moon; \texttt{PyTranSpot} \citep{juvan18}, circular starspots projected on a stellar surface; and the Universal Transit Simulator \citep{deeg09}, a planet with moons or rings.

Meanwhile, significant advances have been made in another problem closely related to shadow imaging: eclipse mapping, which attempts to reconstruct the two-dimensional surface features of an exoplanet undergoing secondary eclipse from the light reflected off the planet's surface as it disappears and reappears from behind the star. \cite{majeau12} and \cite{deWit12} were the first to demonstrate this method, on hot Jupiter HD189733b. \cite{kawahara11} extended this theory to surface mapping of exoplanets in face-on orbits using scattered light, and \cite{farr18} recently released the \texttt{exocartographer} code to carry out surface mapping in a fully Bayesian framework with robust uncertainty estimation. \cite{berdyugina17} showed that next-generation coronagraphic telescopes will be able, using these techniques, to map the surface of a handful of nearby planets, including Proxima b. 

Analogous two-dimensional mapping methods have been successfully applied to the problem of starspot inversion, or deducing the pattern of starspots responsible for time variations in the spectrum or light curve of a star. \cite{goncharskij82} were among the first to attempt starspot inversion, aiming to explain spectral variations in Ap stars by inferring the pattern of chemical inhomogeneities on the surface that would generate them. \cite{vogt83} introduced Doppler imaging to infer maps of starspots on rapidly rotating stars from time series spectra, and \cite{vogt87} refined the technique by introducing maximum entropy regularization as a means of choosing from a set of degenerate solutions to the same observations. \cite{piskunov90} compared the maximum entropy method, which prefers a solution with the minimum spatial correlation between points on the star's surface, to an alternative constraint, Tikhonov regularization, which prefers the smoothest possible pattern of starspots that matches the observations. Similar techniques, with varying choices of regularization, have been applied to stellar light curves by e.g. \cite{lanza98}.

In this work, we build upon these techniques to develop a mathematical and numerical treatment of shadow imaging, which has a similar geometric setup to and is subject to similar degeneracies as eclipse mapping and starspot inversion. In Section~\ref{sec:degeneracies}, we investigate, analytically, the degeneracies inherent to the light curve imaging problem. We explain how discretizing the problem---modeling the transiting object as a grid of pixels of fixed opacity, rather than as a smooth, continuous image---allows us to make progress on the problem despite these degeneracies. In Section~\ref{sec:model}, we define the pixel-grid model which can be used to represent any transiting object and explain how to calculate its light curve. In Section~\ref{sec:fitting}, we consider how, starting from a transit light curve, we may infer the pixel grid image which generated it, and we discuss the results of this inference on a number of test cases. In Section~\ref{sec:realData}, we consider the results of light curve inversion on the real cases of the TRAPPIST-1c,e,f triple transit and the anomalous transits observed in Boyajian's Star. We conclude in Section~\ref{sec:conclusions}.

\section{TRANSIT DEGENERACIES}
\label{sec:degeneracies}

Calculating the light curve of a transiting object is an act of projection. It begins with a three-dimensional object in space, projected against the sky to make a two-dimensional image. At a few discrete points in time, as this image crosses a star, the starlight that the image does not block is summed up, and the sums strung together to make a light curve: a one-dimensional time series.

Deducing the image that generated a particular light curve, therefore, is a problem of inferring two-dimensional data from one-dimensional. As such, we do not expect to find a unique solution to match each light curve. \cite{vogt87, piskunov90, majeau12}, and \cite{deWit12} note similar degeneracies in starspot inversion and eclipse mapping, respectively. 

We begin by examining mathematically the degeneracies inherent to the problem of inferring the shape that generated a particular light curve. We operate under the assumptions, discussed further in~\ref{sec:model}, that the occulting shape is unchanging in time and moving at a constant velocity across the star; that the star is spherical and of uniform brightness; and that the observed light curve is well sampled in time.

\subsection{The Flip Degeneracy}
\label{subsec:flipDegeneracy}

The first important degeneracy in the shadow imaging problem results from the reflection symmetry of the star about its horizontal midplane. An opaque shape that transits at an impact parameter $b$ above the midplane produces the same light curve as a ``flipped" shape that transits below the midplane.

In planetary transit modeling, this degeneracy can be ignored, because the sign of the impact parameter $b = \cos{i}\left(\frac{a}{R_*}\right)\left(\frac{1-e^2}{1+e\sin{\omega}}\right)$ is a function of the inclination angle $i$ of the planet's orbital plane and does not describe any inherent property of the transiting planet. However, if we wish to model more general transiting shapes, we must consider the full space of flip-degenerate solutions.
 
\begin{figure}
\begin{center}
\includegraphics[width=0.45\textwidth]{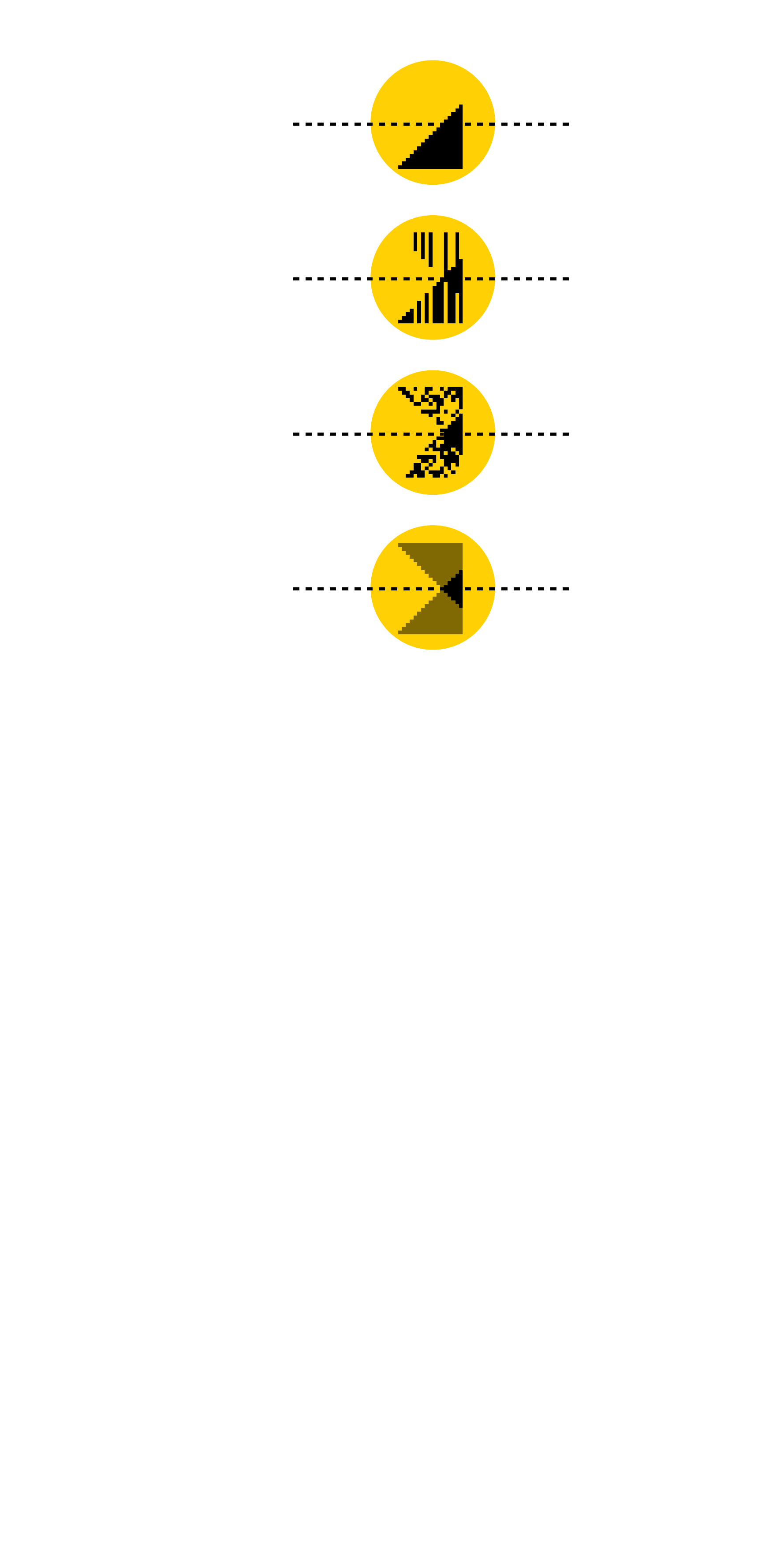}
\caption{Four transiting binary-opacity pixel images which generate the same light curve. The bottom pixel image (opaque black pixels have $\tau=1$; semi-transparent gray pixels have $\tau=0.5$) is the average of the full set of flip-degenerate solutions which match this image's light curve.}
\label{fig:flipDegeneracy}
\end{center}
\end{figure}

To express the degree of flip degeneracy in a given shadow imaging problem, we consider an image made up by a grid of opaque and transparent pixels, $N$ rows by $M$ columns. (See Figure~\ref{fig:flipDegeneracy} for an example.) Although there are $2^{N M}$ unique permutations of opaque and transparent pixels arranged in this grid, each of these permutations does not yield a unique light curve, and in general, a light curve cannot be inverted to produce a unique pixel grid shadow image.

We can express the degree of degeneracy by calculating the number of unique light curves, $U_{LC}$, possible for this $N$-row by $M$-column grid,

\begin{equation}
    U_{LC}= 
\begin{cases}
    \left(2 \times 3^{\frac{(N-1)}{2}}\right)^M ,& \text{$N$ odd } \\
    \left(3^{\frac{N}{2}}\right)^M,              & \text{$N$ even.}
\end{cases}
\label{eq:flip}
\end{equation}

For intuition, consider first the even-$N$ case. In each of the $M$ columns, there are $\frac{N}{2}$ pixels above the midplane; each of these has a counterpart below the midplane with the same impact parameter. There are four possible opacity states of this pair of pixels: $00$, $01$, $10$, or $11$. However, because of the flip degeneracy, the $01$ and $10$ cases produce the same light curve, so only three arrangements are unique. Hence, three unique opacity values for each degenerate pixel pair, raised to the power of the total number of pixels above the midplane.

The arithmetic in the odd-$N$ case is the same, except that the middle pixel row has no across-midplane counterpart. Each pixel in that row may only take on opacity 0 or 1.

In the case of a square, $3 \times 3$ pixel grid, then, there are $2^9 = 512$ unique permutations of opaque and transparent pixels, but only 216 unique light curves. 

As a result, the binary-opacity pixel grid solution to any given light curve inversion is not unique, \textit{unless} the light curve was, in truth, generated by a grid of binary-opacity pixels ($\tau = 0$ or 1) that is symmetrical about its horizontal mid-plane. Physically, we would only expect such a situation for the case of a perfectly spherical planet, or perhaps a planet-moon system or ringed planet, transiting a star at an impact parameter of 0.

In general, therefore, the inverted pixel grid which generates a light curve is not unique. Figure~\ref{fig:flipDegeneracy} shows four transiting pixel images which generate identical light curves. Starting with the pixel image at the top and flipping any pixel about the horizontal mid-plane leaves the light curve unchanged. The pixel image in the bottom panel is the average of the full set of flip-degenerate solutions.

We hope, therefore, to recover shadow images analogous to this bottom panel, which represent a kind of ``superposition" of the full set of flip-degenerate solutions to a particular light curve.

\subsection{The Arc Degeneracy}
\label{subsec:arcDegeneracy}

There is, however, another degeneracy inherent to the shadow imaging problem by which the set of physically allowable images matching any particular light curve becomes infinitely large. This degeneracy allows a transiting pair of semicircular arcs to generate the same light curve as a single opaque point, and we term it the ``arc" degeneracy.

\begin{figure}
\begin{center}
\includegraphics[width=0.45\textwidth]{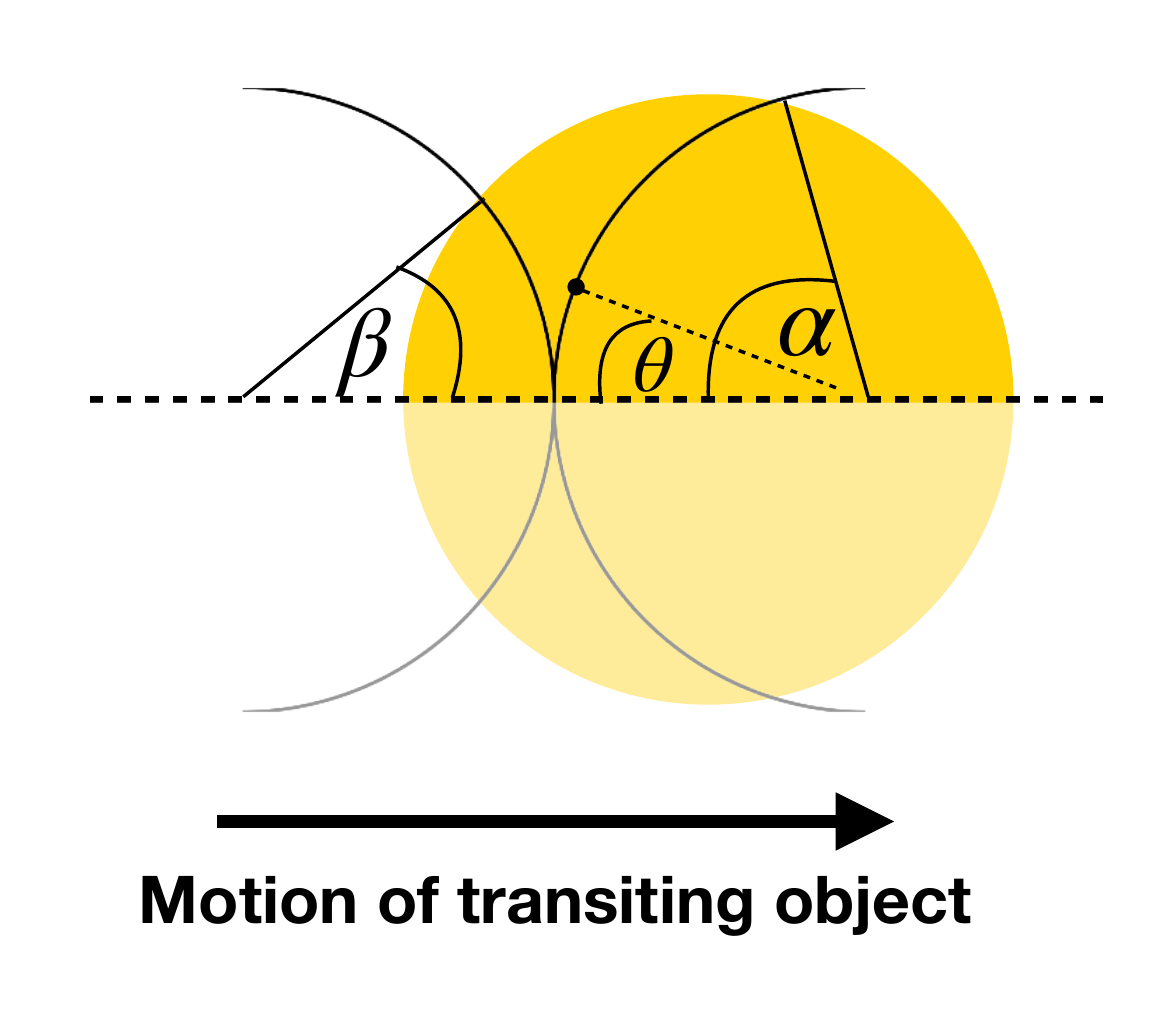}
\caption{A pair of arcs which generates the same light curve as a single opaque point transiting along the horizontal midplane of the star. For this shape to generate a perfect box-like transit, the arcs must be infinitely thin and cannot be of uniform opacity; rather, opacity must be distributed symmetrically along them as a function of $\theta$.}
\label{fig:analyticArcs_angles}
\end{center}
\end{figure}

Figure~\ref{fig:analyticArcs_angles} illustrates the geometry of the pair of arcs which generates the same light curve as an infinitesimally small opaque point transiting exactly along the horizontal midplane of the star. Consider this shape to transit from left to right across the star: because the right-hand arc traces the shape of the stellar limb, the entire right-hand arc will ingress upon the star at the same moment, yielding the same vertical ingress feature in the light curve that we would expect from an infinitesimally small transiting planet. (A correspondingly sharp egress feature in the light curve happens when the left-hand arc egresses all at once some time later.)

After the moment of ingress, the top- and bottom-most edges of the right-hand arc immediately egress again. However, this egress is balanced by the ingress of the middle of the left-hand arc. If opacity is appropriately distributed along each arc, then the ingress of the left-hand arc and egress of the right-hand arc may balance exactly. Here, we derive the functional form of the opacity distribution along the arc to allow this exact balance.

Let $\alpha$ (see Figure~\ref{fig:analyticArcs_angles}) denote the angle between the horizontal midplane of the star and the point of intersection between the stellar limb and the right-hand arc (which ingresses first). At the moment of ingress, $\alpha = \frac{\pi}{2}$; at the moment of egress, $\alpha = 0$. Let $\beta$ denote the corresponding angle to the point of intersection on the left-hand arc, and let $\beta$ range from $0$ at ingress to $\frac{\pi}{2}$ at egress.

\begin{figure}
\begin{center}
\includegraphics[width=0.45\textwidth]{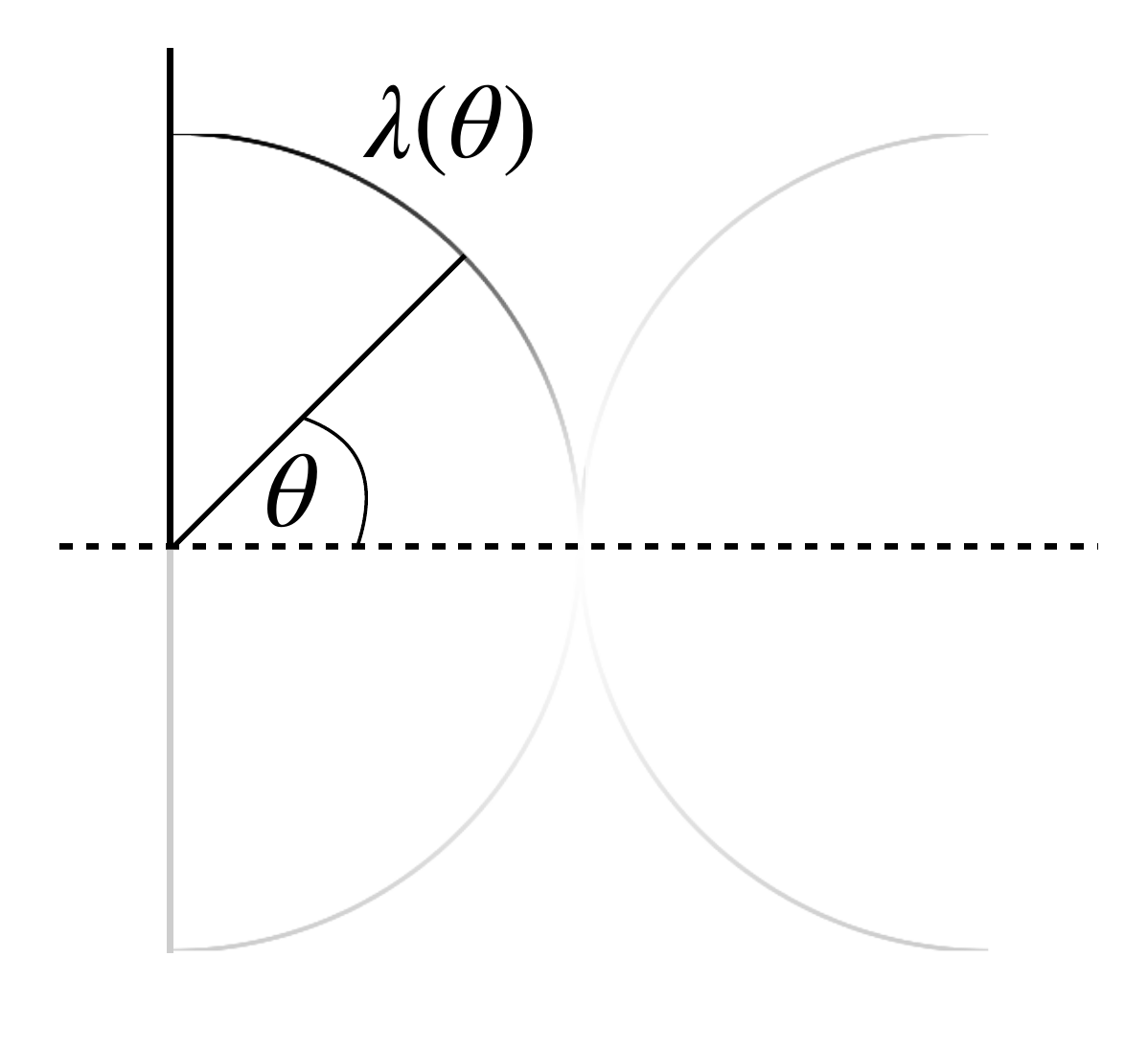}
\caption{$\theta$ represents the angle from the horizontal midplane of either arc to any point along it. $\lambda(\theta)$ represents the opacity of the arc at $\theta$. We wish to solve for $\lambda(\theta)$ such that the arc pair can produce a flat-bottomed transit.}
\label{fig:analyticArcs_lambda}
\end{center}
\end{figure}

Let $\theta$ represent an angle measured from the horizontal midplane of either arc to its outermost point, and let $\lambda(\theta)$ represent the opacity along the arc as a function of this angle. Figure~\ref{fig:analyticArcs_lambda} illustrates this setup. Note that $\lambda(\theta)$ cannot be constant, because, for example, during some small time interval $dt$ immediately after the moment of ingress, the length of the right-hand arc which egresses is greater than the length of left-hand arc which ingresses. 

Let $T$ be the duration of the transit of the pair of arcs (in other words, the interval between the moment of ingress and the moment of egress). Let the moment of ingress happen at $t = 0$, and let us define a dimensionless time coordinate $\kappa = \frac{t}{T}$ to parametrize the progress of the transit. At ingress, then, $\kappa = 0$, and at egress, $\kappa = 1$.

Following these definitions, we may write

\begin{equation}
    \cos{\alpha} = \kappa,
\end{equation}

\begin{equation}
    \cos{\beta} = 1 - \kappa.
\end{equation}

The total opacity $L(\kappa)$ transiting the star at a particular moment $\kappa$ is equal to

\begin{equation}
    L(\kappa) = \int_{0}^{\alpha(\kappa)} \lambda(\theta) d\theta + \int_{0}^{\beta(\kappa)} \lambda(\theta) d\theta.
\label{eq:transitingOpacity}
\end{equation}

For the transit to be flat-bottomed, we require that $L(
\kappa)$ be constant, or that $\frac{dL}{d\kappa} = 0$. Differentiating both sides of Equation~(\ref{eq:transitingOpacity}) by $\kappa$, we obtain

\begin{equation}
    \frac{dL}{d\kappa} = \lambda(\alpha) \frac{d\alpha}{d\kappa} + \lambda(\beta) \frac{d\beta}{d\kappa},
\end{equation}

because $\lambda$ is time-independent and therefore independent of $\kappa$. 

Setting this expression equal to 0 and substituting, we obtain

\begin{equation}
    0 = \lambda(\alpha) \frac{1}{\sqrt{1 - \kappa^2}} + \lambda(\beta) \frac{1}{\sqrt{1 - (1 - \kappa)^2}},
\label{eq:setdLdkequalto0}
\end{equation}

or

\begin{equation}
    \frac{\lambda(\alpha)}{\lambda(\beta)} =  -\frac{\sqrt{1 - \kappa^2}}{\sqrt{1 - (1 - \kappa)^2}}.
\end{equation}

By the definitions of $\alpha$ and $\beta$, we may write

\begin{equation}
    \frac{\lambda(\alpha)}{\lambda(\beta)} =  -\frac{\sqrt{1 - \cos^2{\alpha}}}{\sqrt{1 - 
    \cos^2{\beta}}} = \frac{\sin{\alpha}}{\sin{\beta}}.
\end{equation}

By inspection, then, 
\begin{equation}
    \lambda(\theta) \propto \sin{\theta},
\label{eq:sineSoln}
\end{equation}

where we choose the sign to be positive because physically meaningful opacities are between 0 and 1. 

\begin{figure}
\begin{center}
\includegraphics[width=0.45\textwidth]{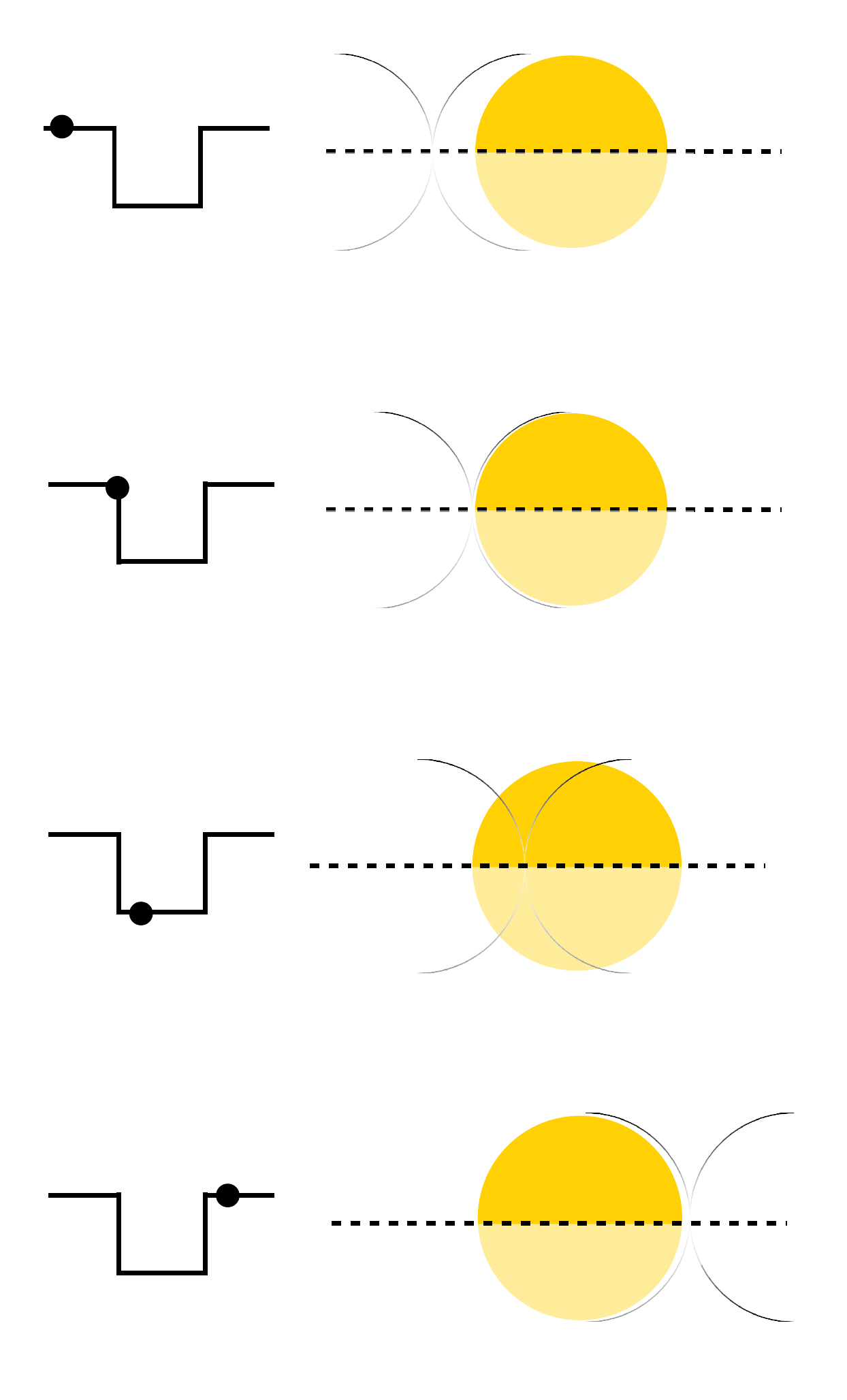}
\caption{A transiting arc pair with opacity distributed as $\lambda(\theta) = \sin{\theta}$. This shape generates a box-like transit light curve. The circles in the left-hand panels mark the time along the transit at which the right-hand panels occur.}
\label{fig:analyticArcs_transit}
\end{center}
\end{figure}

The overall normalization of $\lambda(\theta)$ sets the transit depth of the arcs' light curve. Figure~\ref{fig:analyticArcs_transit} shows a transiting arc pair with $\lambda(\theta) = \sin{\theta}$.

We note that there are two other solutions to $\lambda(\theta)$ that satisfy the condition that $\frac{dL}{d\kappa} = 0$. The first is the trivial solution, $\lambda(\theta) = 0$. The second is a Dirac delta function at $\theta = 0$, 

\begin{equation}
    \lambda(\theta) \propto \delta(\theta = 0),
\label{eq:deltaSoln}
\end{equation}

where again the overall normalization sets the transit depth.

For intuition, the two non-trivial solutions to $\lambda(\theta)$ given by Equations~(\ref{eq:sineSoln}) and~(\ref{eq:deltaSoln}) represent two extremes: the least and most compact arrangements of opacity, respectively, that produce the same flat-bottomed, box-like transit. Any linear combination of these solutions also satisfies $\frac{dL}{d\kappa} = 0$ and generates a box-like transit.

\begin{figure}
\begin{center}
\includegraphics[width=0.3\textwidth]{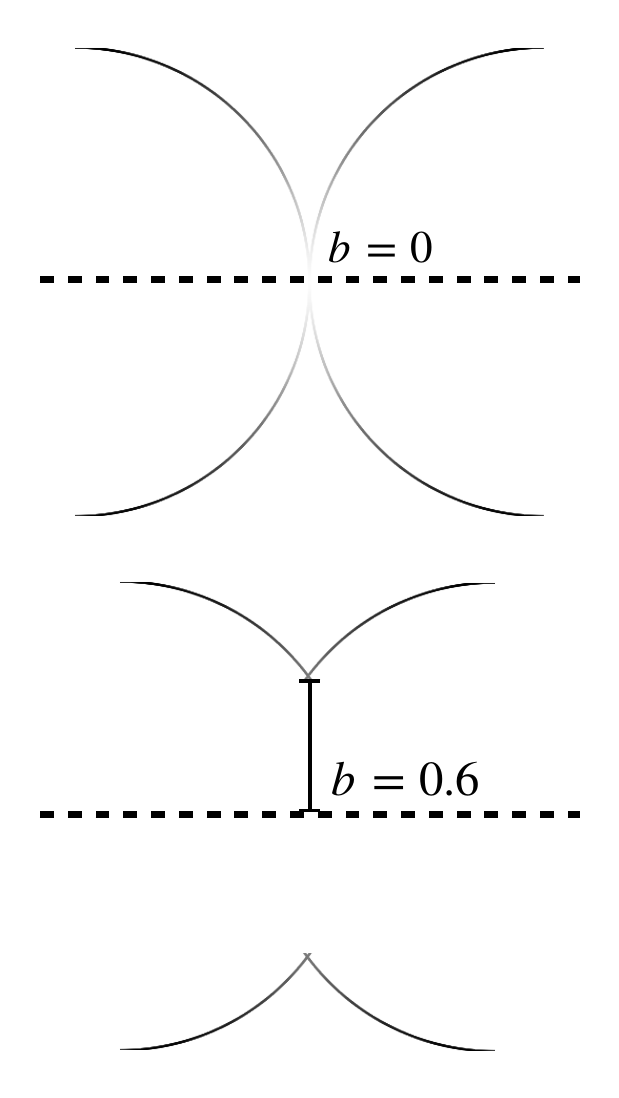}
\caption{A pair of truncated arcs, as illustrated in the lower panel, can match the transit shape of an infinitesimal opaque point transiting at arbitrary impact parameter.}
\label{fig:analyticArcs_b}
\end{center}
\end{figure}

The above derivation has demonstrated that a pair of arcs of variable opacity can match the transit shape of an infinitesimal point of opacity transiting along the horizontal midplane of the star, at impact parameter $b = 0$. The same logic applies to an infinitesimal point at arbitrary impact parameter $b$. Figure~\ref{fig:analyticArcs_b} illustrates the geometry of this situation.

Mathematically, a change in the impact parameter $b$ means that the limits of integration in Equation~(\ref{eq:transitingOpacity}) change,

\begin{equation}
    L(\kappa) = \int_{\arcsin{b}}^{\alpha(\kappa)} \lambda(\theta) d\theta + \int_{\arcsin{b}}^{\beta(\kappa)} \lambda(\theta) d\theta.
\label{eq:transitingOpacity_b}
\end{equation}

Since $b$ is constant, the subsequent steps and resulting solutions for $\lambda(\theta)$ do not change, except that the delta function solution is localized at $\theta = \arcsin{b}$.

We note finally that the arc degeneracy technically only operates for an occulter transiting a uniformly bright star: if the star is limb-darkened, then there is no (unchanging) arc arrangement which can maintain the perfect opacity ingress-egress balance described by Equation~(\ref{eq:setdLdkequalto0}). However, in practice, the limited time resolution of light curve observations leaves room for significant arc-degenerate behavior in shadow images recovered from real transit data (see~\ref{sec:realData} below).

\subsection{The Stretch Degeneracy}
\label{subsec:wvDegeneracy}
A third degeneracy inherent to light curve imaging results from the ``scale-free" nature of the problem, and allows a wide image moving at high velocity to generate the same light curve, within an arbitrarily small measurement uncertainty, as a narrower image moving at lower velocity. We term this degeneracy the ``stretch" degeneracy.

The stretch degeneracy is mathematically simpler than the arc or flip degeneracies. Two occulters with the same transit duration $T$ both obey

\begin{equation}
    T = \frac{W}{v},
\label{eq:wvDegeneracy}
\end{equation}

where $W$ is the width of the occulter, and $v$ is its velocity. The right-hand side of this equation can be multiplied by the same constant in the numerator and denominator without consequence to $T$. In other words, a ``stretched" image traveling fast can generate a transit event of the same duration as a narrow image traveling slowly.

\begin{figure}
\begin{center}
\includegraphics[width=0.45\textwidth]{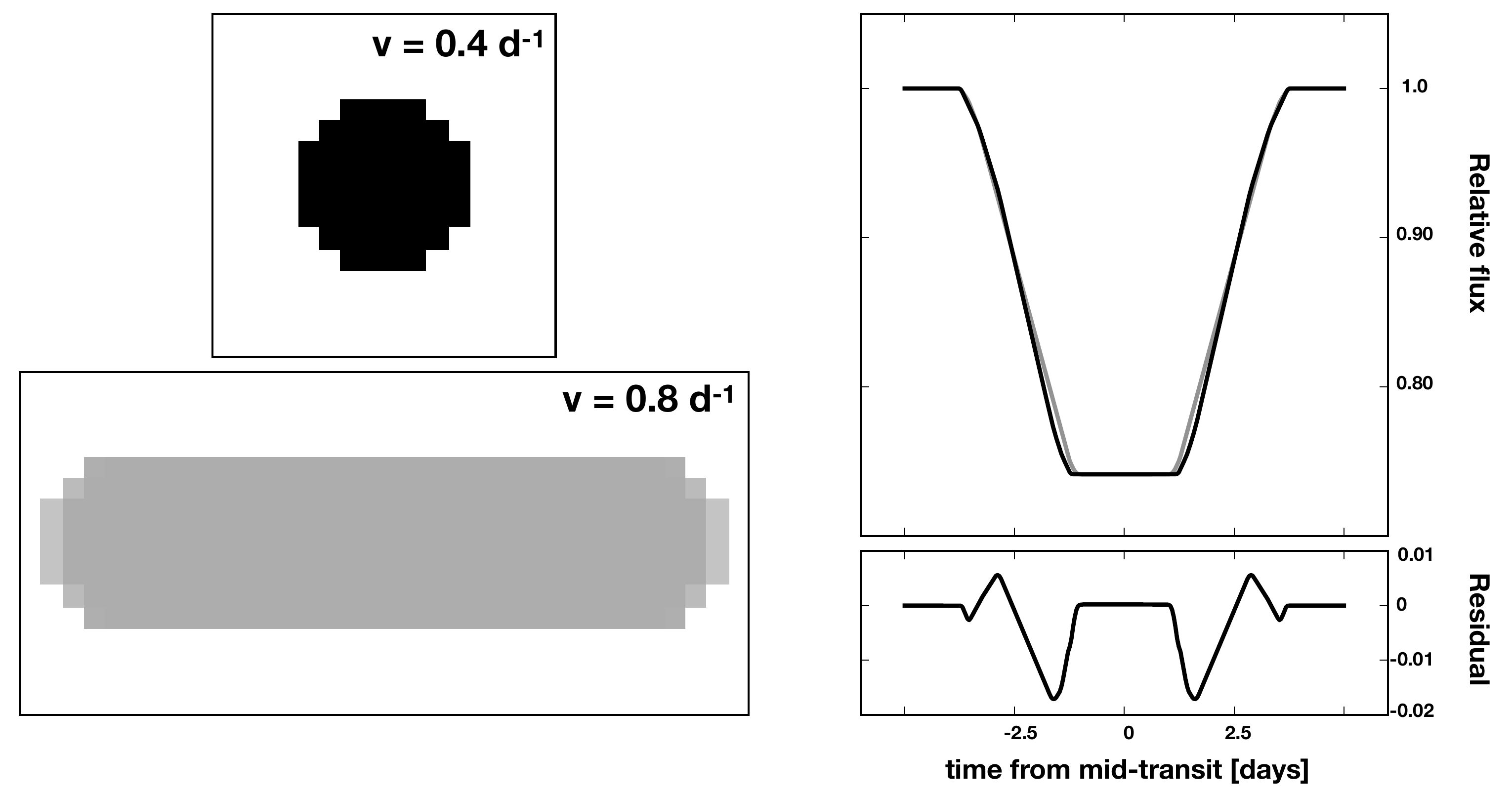}
\caption{Two transiting images with light curves that differ by $\mathcal{O}(1\%)$. The lower image transits at a velocity twice that of the upper image. Note that the left- and rightmost edges of the lower image are slightly less opaque than the middle, an adjustment made to better match the ingress and egress shape of the upper image's light curve. At higher resolution for the lower image, an even better match to the upper image's light curve could be found.}
\label{fig:wvDegeneracy}
\end{center}
\end{figure}

Figure~\ref{fig:wvDegeneracy} illustrates the stretch degeneracy for a simple, low-resolution circular occulter. Note in particular two features of the ``stretched" image: first, that it is semi-opaque rather than fully opaque like the un-stretched image, and second, that its edges are less opaque than its middle. The semi-opacity of the stretched image is necessary in order to match the transit depth of the un-stretched image: because the stretched image is wider, it occults more of the stellar surface, so it must let some light through, or it will produce a much deeper transit than the un-stretched image. Meanwhile, the lightened edges of the stretched image are necessary to better match the ingress and egress shape of the un-stretched image's light curve; with arbitrarily high image resolution, it is possible to match the un-stretched image's light curve to arbitrary accuracy.

In practice, the stretch degeneracy is the least important of the three non-trivial degeneracies we explore in this section, because a fast-transiting, stretch-degenerate image can only match a narrow, slower image's light curve if the image resolution is high enough, as suggested by the example in Figure~\ref{fig:wvDegeneracy}. For real data, image resolution is constrained by the number of observed data points, which causes us to prefer the narrowest, slowest possible image which can match an observed light curve (see~\ref{subsec:gridParams} for further discussion).

\subsection{Trivial Degeneracies}
\label{subsec:trivialDegeneracy}

Finally, we note two trivial degeneracies which do not affect the inference of a shadow image. The first relates to the arbitrary sign of the velocity of the transiter; an image which transits left-to-right across the star generates the same light curve as the same image, horizontally mirrored, transiting right-to-left across the star at the same velocity. We choose positive $v$ to indicate that the image transits left-to-right (see~\ref{subsec:gridDef}, below).

The second trivial degeneracy relates to a time translation of the entire transit event. As we discuss in~\ref{subsec:gridDef}, we must choose a ``reference time," analogous to a transit midpoint time, along a light curve in order to recover a shadow image; shifting this reference time forward or backward along the light curve results in a shadow image which is shifted right or left, respectively (given our choice of $v$ direction, above).

\section{MODEL: GENERATING A LIGHT CURVE FROM A DISCRETIZED IMAGE}
\label{sec:model}

By the arguments of Section~\ref{sec:degeneracies}, a given light curve may be generated by infinitely many images. To constrain the solution set, we therefore conclude that it is necessary to impose further constraints on the shadow image. (Starspot inversion requires an analogous constraint---popular choices include the maximum entropy principle, which chooses the solution with minimum spatial correlation between points on the stellar surface, and Tikhonov regularization, which chooses the smoothest solution, or the solution with minimum spatial derivative.)

In this section, we define a forward model for generating a light curve, sampled at discrete time intervals, from a pixelated image. This simulated light curve can be compared to observations of a real transit event. After we establish this forward model, we investigate the inverse problem, of how to infer a pixelated image from an observed light curve, in the next section. We return to the question of degeneracies in~\ref{subsec:pixelOpacities}.


\subsection{Discretizing the Image}\label{subsec:discretization}




Pixelating, or discretizing, the shadow image is motivated by recognizing that real light curves are themselves discrete time series. A light curve is not infinitely resolved in time, and therefore we should not attempt to recover a shadow image that is infinitely resolved spatially. Similarly, each flux measurement in a light curve has an associated uncertainty; we should not attempt to recover a shadow image with pixel elements too small to be definitively detected within that uncertainty (see Section~\ref{subsec:gridParams} below for further discussion).

Discretizing the pixel image, furthermore, enables us to investigate two physical variants of the shadow imaging problem:
\begin{enumerate}
    \item What if the transiting object which generated the light curve is a solid body, and therefore our shadow image should only admit of completely transparent (opacity $\tau = 0$) or completely opaque ($\tau$ = 1) pixels?
    \item What if the transiting object is dusty or translucent, or is a solid body smaller than the pixel scale, and our shadow image can contain pixels of intermediate opacity ($0 \leq \tau \leq 1$)?
\end{enumerate}

These two variations of the shadow imaging problem have different constraints on the pixel opacities, and require different mathematical approaches to inversion. In case (1), discretizing the pixel image is necessary to divide it up into opaque and transparent elements. In case (2), discretizing the pixel image enables us to set up the light curve inversion problem as a single matrix equation (Equation~(\ref{eq:nonsquarematrixform_simple}), below), and to explore both analytic and numerical approaches to solving this equation. (Similar mathematical formulations exist for both starspot inversion \citep{vogt87} and eclipse mapping, e.g. \citealt{berdyugina17}.)

The same forward model, or procedure for generating a light curve from a pixelated image, can be used in both cases, so we begin there. How do we calculate the light curve of a pixelated image grid transiting a star?

\subsection{Grid Definitions and Positions}
\label{subsec:gridDef}

We consider a pixel grid of $N$ rows and $M$ columns transiting a star. We normalize the physical scale of the problem such that the radius of the star is unity. 

\begin{figure*}
\begin{center}
\begin{overpic}[width=18.0cm,angle=0,clip=true]{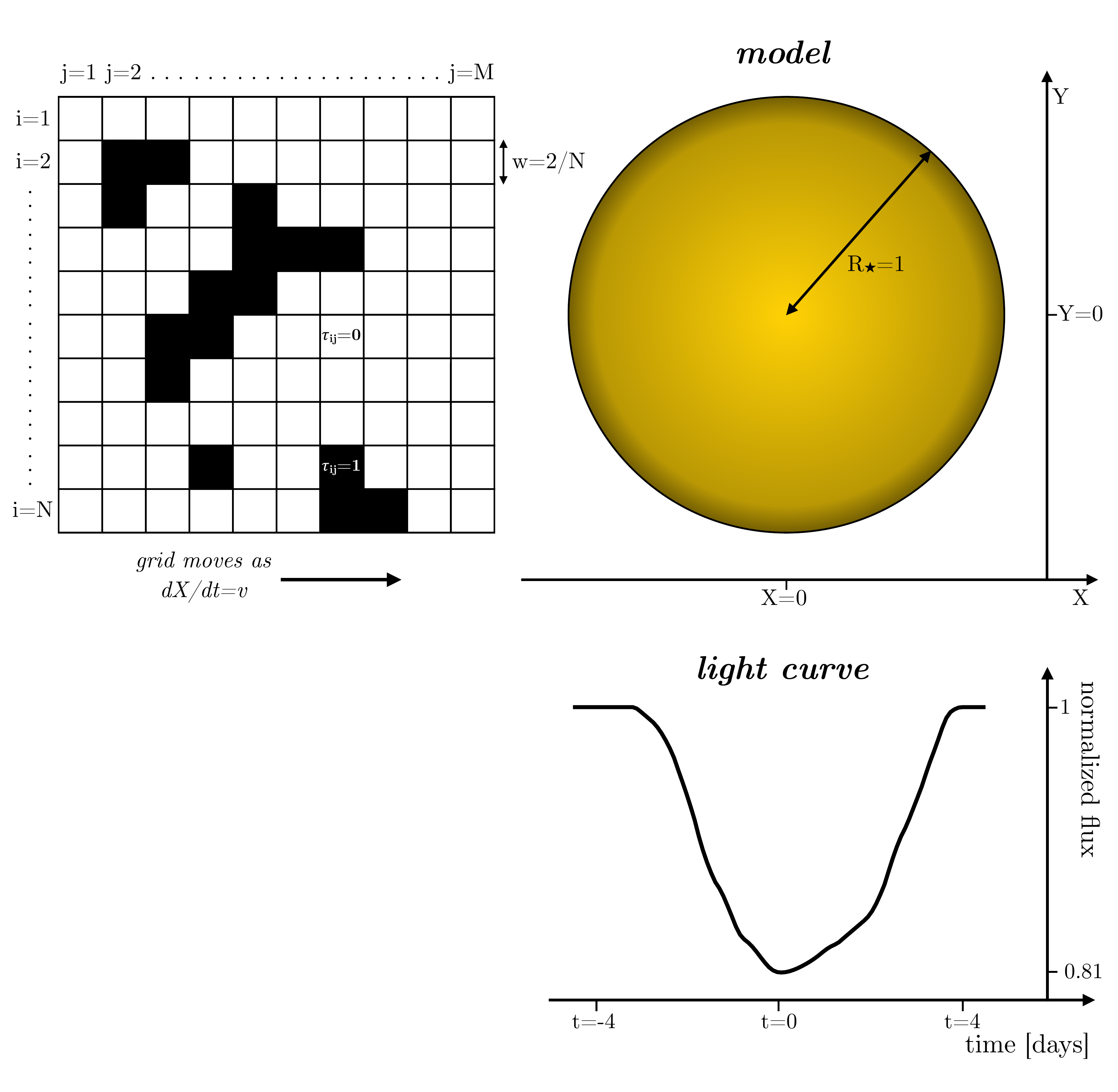}

\put (5,20) {{\parbox{0.39\linewidth}{
\textbf{Figure 7.} (Top panels) Illustration of a $N=10$ by $M=10$ binary-opacity grid model with 16 opaque pixels. The star itself is not pixelated; rather, the pixelated grid transits across the star and the exact area of overlap of each square pixel and the star is evaluated at each discrete time step in order to generate a light curve. (Bottom panel) The light curve generated when this grid transits across a uniformly bright star at $v = 0.4\  \mathrm{days}^{-1},\ t_{\mathrm{ref}} = 0\  \mathrm{days}$.
}}}
\end{overpic}
\caption{}
\label{fig:setup}
\end{center}
\end{figure*}

The grid lives in the $X$-$Y$ sky-projected plane, with the observer at $Z=+\infty$. The grid moves laterally along the $X$ axis, with $\mathrm{d}X/\mathrm{d}t>0$, and does not translate up or down (i.e. $\mathrm{d}Y/\mathrm{d}t=0$). We illustrate this setup in Figure~\ref{fig:setup}.

We treat the grid as moving at a constant lateral velocity $v \equiv \mathrm{d}X/\mathrm{d}t$, where $\mathrm{d}v/\mathrm{d}t\equiv0$. This is a reasonable approximation over the timescale of a transit, unless the object resides on a very tight orbit, or the object is near pericenter on a highly eccentric orbit. We define positive $v$ to mean that the grid transits from left to right across the star, such that the rightmost column of pixels ingresses first.

We define the vertical position of the grid such that the top of the highest row of pixels falls at $Y=1$ and the
bottom of the lowest row of pixels falls at $Y=-1$. In this way, the grid perfectly overlaps with the star in the vertical dimension. 

This definition sets the size of each pixel to have a width, $w$, of

\begin{align}
w = 2/N.
\label{eq:pixelwidth}
\end{align}

We emphasize that every pixel has the same square shape with this dimension. For $N=1$, then, $w=2$ and is thus equal to the diameter of the star. 

To refer to individual pixels, we adopt the index notation $i \in [1, N]$ to denote the row and $j \in [1, M]$ to denote the
column. To calculate the amount of stellar flux the grid blocks at each discrete time step $t_k$ of the transit observation, we must first calculate the $X$ and $Y$ positions of each grid pixel $i,j$ at each time step.

We may write the $Y$-position of the center of pixel $i, j$ as

\begin{align}
Y_{i,j} &= 1 - (w/2) - (i-1) w,
\label{eq:Ypos}
\end{align}

where setting $i=1$ recovers $Y_{1,j}=1-(w/2)$, and setting $i=N$ recovers $Y_{N,j}=-1+(w/2)$. The $Y$ positions of the grid pixels are constant.

For the $X$ positions of the pixels, which evolve in time, we first define a reference $X$
position for each pixel at a reference time $t=t_{\mathrm{ref}}$ as

\begin{align}
X_{i,j}^{\mathrm{ref}} &\equiv X_{i,j}[t=t_{\mathrm{ref}}].
\end{align}

We choose the reference time such that the grid is centered on the star at $t=t_{\mathrm{ref}}$. Therefore:

\begin{align}
X_{i,j}^{\mathrm{ref}} &= (j-j_{\mathrm{mid}}) w,
\end{align}

where

\begin{align}
j_{\mathrm{mid}} &= 1 + (M-1)/2.
\end{align}

We may now write the time-evolving $X$-position of the center of any pixel as

\begin{align}
X_{i,j}(t_k) &= X_{i,j}^{\mathrm{ref}} + (t_k - t_{\mathrm{ref}}) v,
\label{eq:Xpos}
\end{align}

where $t_k$ is the $k^{\mathrm{th}}$ time index, and $k \in [1, K]$. Practically speaking,
$t_{\mathrm{ref}}$ is analogous to the transit mid-time
fitted in conventional transit models.

We may use the above equation for the time-evolving $X_{i,j}$ to solve for the
time at which the grid makes first and last contact with the star,
$t_{\mathrm{enter}}$ and $t_{\mathrm{exit}}$. The grid moves from left to right across the star, so at first contact, the
$M^{\mathrm{th}}$ column of pixels has an $X$ position equal to $-1-(w/2)$,
and for the last contact the $1^{\mathrm{st}}$ column of pixels has an
$X$ position equal to $1+(w/2)$, giving

\begin{align}
t_{\mathrm{enter}} &= t_{\mathrm{ref}} - \frac{1 + M w/2}{v},\\
t_{\mathrm{exit}}  &= t_{\mathrm{ref}} + \frac{1 + M w/2}{v}.
\label{eq:tBoundaries}
\end{align}

We assign a time-independent opacity $\tau_{i,j}$ to each pixel. $\tau_{i,j}$ is a binary value equal to zero or unity---in other words, we construct our grid of perfectly transparent pixels ($\tau_{i,j} = 0$) and opaque pixels ($\tau_{i,j} = 1$). 

In total then, our model has $M N$ opacity parameters, which are binary-valued (case 1) or real numbers between 0 and 1, inclusive (case 2), and two auxiliary parameters, $t_{\mathrm{ref}}$ and $v$, which are real-valued.

\subsection{Computing the Light Curve of a Pixel}

As pixel $i, j$ transits the star, it occludes a fractional area $A_{i,j}(t_k)$ of the stellar disk at time $t_k$; $A = 0$ for pixels which do not overlap the star, and $A = \frac{w^2}{\pi}$ for pixels which overlap completely (since we choose $R = 1$, the area of the entire stellar disk is equal to $\pi$). 

If we assume that the stellar disk is uniformly bright (i.e., there is no limb darkening), we may then compute the light curve $\mathbfit{F}(t)$ of the transiting grid by recognizing that the fractional flux \textit{blocked} by the grid at each time step $t_k$ is equal to the fractional area of the star occulted by non-transparent pixels ($\tau_{i,j} > 0$), in proportion to their opacity. Therefore, the \textit{unocculted} flux at time $t_k$ is given by:

\begin{align}
F(t_k) &= 1 - \sum_{i=1}^N \sum_{j=1}^M \tau_{i,j} A_{i,j}(t_k).
\label{eq:LCuniform}
\end{align}

This is the equation for the transit light curve, normalized such that $F = 1$ out-of-transit.

We emphasize that, while opacities $\tau < 0$ and $\tau > 1$ are mathematically permissible in this equation, they are unphysical: a $\tau < 0$ would represent a transiting pixel brighter than the stellar surface it occulted, and a $\tau > 0$ would describe a pixel that blocked more than its proper area's worth of starlight.

We compute the area of overlap $A_{i,j}(t_k)$ of pixel $i, j$ at time step $t_k$ from the $(X, Y)$ position of the pixel's center at $t_k$, given by Equations~(\ref{eq:Ypos}) and~(\ref{eq:Xpos}), and the pixel's width, given by Equation~(\ref{eq:pixelwidth}). When a pixel partially overlaps the star, we approximate its overlap area as either a triangle, a trapezoid, or a square missing a triangular corner. We choose the appropriate overlap-shape by computing the number of intersection points between the edge of the star and the sides of the pixel, and also noting whether the center of the pixel falls inside or outside the star. We then correct this approximation by using the length of the chord between intersection points to calculate the area of the sliver of occluded star yet unaccounted for by this approximation.

For a limb-darkened star, we must also account for the position of each opaque pixel relative to the stellar limb at each time step in order to determine how much flux it occludes. We adopt the small-planet approximation of \cite{mandelagol}, in which it is assumed that the star's surface brightness is constant across a pixel. In other words, we treat the pixel as occulting a thin, uniform-surface-brightness, annular slice of the stellar disk, where the radius of the annulus is the distance from the center of the stellar disk to the center of the pixel, and the annulus is just wide enough to encompass the pixel. 

We denote the area of this annulus as $A_{\mathrm{annulus}}$, and its emitted flux as $F_{\mathrm{annulus}}$. As a rule of thumb, this small-planet approximation is only appropriate for $w \lesssim 0.2$ (i.e. $N > 10$), which corresponds roughly to an occulter-to-star ratio-of-radii of $0.1$, for a circular occulter of the same area as the pixel.  The exact ratio-of-radii at which the small-planet approximation becomes inappropriate depends on the impact parameter of the pixel, the size of the pixel, the limb-darkening profile of the star, and the bandpass of the observations, so there is no general exact cutoff.

To calculate the light curve in the limb-darkened case, we must re-normalize Equation~(\ref{eq:LCuniform}): the fractional flux occulted by an opaque pixel is no longer equal to the fractional area of the stellar disk occluded by the pixel, but rather to:
\begin{equation}
    \bar{A}_{i,j}(t_k) = \frac{A_{i,j}(t_k)}{A_{\mathrm{annulus}}}  \frac{F_{\mathrm{annulus}}}{F_{\mathrm{\star}}},
\label{eq:overlapareas}
\end{equation}

where $F_{\mathrm{\star}}$ is equal to the flux of the entire unocculted, limb-darkened star, relative to the non-limb-darkened star (which must be calculated given a choice of limb-darkening coefficients). We note that this equation reduces to $\bar{A}_{i,j}(t_k) =A_{i,j}(t_k)$ in the case of uniform limb-darkening.

The value of the light curve at $t_k$ is then given by:

\begin{align}
F(t_k) &= 1 - \sum_{i=1}^N \sum_{j=1}^M \tau_{i,j} \bar{A}_{i,j}(t_k).
\label{eq:LClimbdarkened}
\end{align}

We provide \texttt{Python} code to calculate the transit light curve of any grid in the case of uniform, linear, quadratic, or 4-parameter nonlinear limb-darkening in the software package accompanying this paper, \texttt{EightBitTransit}.

\section{FITTING: SHADOW IMAGING A PIXEL GRID FROM A LIGHT CURVE}
\label{sec:fitting}

In this section, we describe how we use the forward model described above to solve the inverse problem, ``shadow imaging." We observe a light curve $\mathbfit{F}$, made up of discrete flux measurements $F_k \equiv F(t_k)$ over $K$ points in time: what pixelated image generated that light curve?

To illustrate the complexity of this problem, we begin with an order-of-magnitude estimation of the number of arrangements of pixels in a binary-valued shadow image (case (1)). There are $2^{N M}$ unique permutations of transparent and opaque pixels for an $N$ by $M$ grid, and $\mathcal{O}[3^{NM/2}]$ unique light curves that can result (by the flip degeneracy, discussed in~\ref{subsec:flipDegeneracy}). For a 10 by 10 grid, then, there are $\mathcal{O}[10^{30}]$ unique permutations of the binary pixel opacities; accounting for the flip degeneracy, if one wished to find the binary pixel opacity arrangement of the just top half of a 10 by 10 grid to best match a particular light curve, one would have to evaluate $\mathcal{O}[10^{24}]$ possibilities. 

A full parameter search is therefore not practically feasible. The largest square grid which could be reasonably fully searched is 5 by 5, for which there are 33.6 million full-grid permutations and 1.9 million half-grid permutations (by Equation~(\ref{eq:flip})). We must therefore infer the pixel opacities from the light curve, not attempt to guess them.

To infer a pixel grid from a light curve $\mathbfit{F}$, we must first select the grid parameters: the dimensions $N$ and $M$, the velocity $v$, and the reference time $t_{\mathrm{ref}}$. Given these choices, we may calculate the areas of overlap of each grid pixel at each light curve time step, and the corresponding $\bar{A}_{i,j}(t_k)$ for any choice of limb-darkening law. All that remains is to solve Equation~(\ref{eq:LClimbdarkened}) for the opacities of the grid pixels, $\tau_{i,j}$, subject to the constraints of either case (1) ($\tau_{i,j} = 0$ or 1) or case (2) ($0 \leq \tau_{i,j} \leq 1$).

\subsection{Mathematical Setup}

To be exact, we note that $\mathbfit{F}$ is a column vector of length $K$, of which each scalar entry $F_k \equiv F(t_k)$ is given by Equation~(\ref{eq:LClimbdarkened}). Let us ``unravel" the double sum in Equation~(\ref{eq:LClimbdarkened}) by defining a new index $l$, such that

\begin{equation}
l[i,j] = j + (i-1) M.
\end{equation}

Since $i$ ranges from 1 to $N$, and $j$ from 1 to $M$, $l$ ranges from 1 to $MN$. 

We may then rewrite Equation~(\ref{eq:LClimbdarkened}) as

\begin{align}
F_k &= 1 - \sum_{l=1}^{L} \tau_{l} \bar{A}_{l}(t_k),
\label{eq:LClimbdarkenedraveled}
\end{align}

where we define $L \equiv MN$. If we further define $\bar{A}_{k,l} \equiv \bar{A}_{l}(t_k)$, then

\begin{align}
F_k &= 1 - \sum_{l=1}^{L} \tau_{l} \bar{A}_{k,l}.
\label{eq:LClimbdarkenedraveledsubscript}
\end{align}

Let us now rewrite $\bar{A}_{k,l}$ in matrix form:

\begin{eqnarray}
 \xoverline{\mathbfss{A}} = 
  \begin{pmatrix}
    a_{1,1} & a_{1,2} & \cdots & a_{1,L} \\
    a_{2,1} & a_{2,2} & \cdots & a_{2,L} \\
    \vdots & \vdots & \ddots & \vdots \\
	a_{K,1} & a_{K,2} & \cdots & a_{K,L}
  \end{pmatrix}.
\label{eqn:thematrix}
\end{eqnarray}

$\xoverline{\mathbfss{A}}$ is a matrix of shape $K$ by $L$, where the $k^{\mathrm{th}}$ row encodes the state of overlap of the entire pixel grid at time step $k$, and the $l^{\mathrm{th}}$ column encodes the overlap state of pixel $l$ across all time steps.

Similarly, we may ``unravel" the opacity matrix $\tau$ into a column vector $\boldsymbol{\tau}$ of length $L$:

\begin{eqnarray}
\boldsymbol{\tau} = 
  \begin{pmatrix}
    \tau_1 \\
    \tau_2 \\
    \vdots \\
	\tau_L 
  \end{pmatrix}.
\end{eqnarray}

We may now re-express Equation~(\ref{eq:LClimbdarkened}) in matrix form:

\begin{equation}
\mathbfit{F} = \mathbfit{1} - \xoverline{\mathbfss{A}}\boldsymbol{\tau},
\end{equation}

where

\begin{eqnarray}
\mathbfit{F} = 
  \begin{pmatrix}
    F_1 \\
    F_2 \\
    \vdots \\
	  F_k
  \end{pmatrix}
\end{eqnarray}

and $\mathbfit{1}$ is a column vector of ones, equal in length to $\mathbfit{F}$.

If we define a vector $\mathbfit{R} = \mathbfit{1} - \mathbfit{F}$, we may rearrange this equation to read

\begin{equation}
\xoverline{\mathbfss{A}}\boldsymbol{\tau} = \mathbfit{R}.
\label{eq:matrixform}
\end{equation}

If $\xoverline{\mathbfss{A}}$ were invertible, then our work would be done: we could solve Equation~(\ref{eq:matrixform}) directly for the vector of pixel opacities $\boldsymbol{\tau}$. However, because of the flip degeneracy, pixel $i, j$ has the same area-of-overlap at every time step as pixel $(N+1-i), j$, and as a result, $\xoverline{\mathbfss{A}}$ always has repeated columns. By the invertible matrix theorem, a matrix with repeated columns is not invertible. 

We may proceed by recognizing that $\xoverline{\mathbfss{A}}$ and $\boldsymbol{\tau}$, since they describe the entire pixel grid, contain redundant information. We need only solve for the opacities of one half of the pixels (we choose the top half, for convenience). We define a new index

\begin{equation}
    L_{\mathrm{half}} 
\begin{cases}
    \frac{(N-1)M}{2} + M ,& \text{$N$ odd } \\
    \frac{NM}{2},              & \text{$N$ even.}
\end{cases}
\end{equation}

We define a new area-of-overlap matrix $\xoverline{\mathbfss{A}}_{\mathrm{half}}$, which represents the left half (columns 1 through $L_{\mathrm{half}}$, inclusive) of $\xoverline{\mathbfss{A}}$, and a new opacity vector $\boldsymbol{\tau}_{\mathrm{half}}$, which represents the corresponding top half of $\boldsymbol{\tau}$. We may then write:

\begin{equation}
\xoverline{\mathbfss{A}}_{\mathrm{half}}\boldsymbol{\tau}_{\mathrm{half}} = \mathbfit{R}.
\label{eq:matrixform_half}
\end{equation}

Since, in general, $K \neq L_{\mathrm{half}}$, we may multiply both sides of this equation by $\xoverline{\mathbfss{A}}_{\mathrm{half}}^{\mathrm{T}}$ to yield

\begin{equation}
\xoverline{\mathbfss{A}}_{\mathrm{half}}^{\mathrm{T}}\xoverline{\mathbfss{A}}_{\mathrm{half}}\boldsymbol{\tau}_{\mathrm{half}} = \xoverline{\mathbfss{A}}_{\mathrm{half}}^{\mathrm{T}}\mathbfit{R}
\label{eq:nonsquarematrixform}
\end{equation}

so that both sides of the equation are column vectors of length $L_{\mathrm{half}}$, and $\xoverline{\mathbfss{A}}_{\mathrm{half}}^{\mathrm{T}}\xoverline{\mathbfss{A}}_{\mathrm{half}}$ is a square matrix. 

For notational simplicity, let $\mathbfss{B} \equiv \xoverline{\mathbfss{A}}_{\mathrm{half}}^\mathrm{T}\xoverline{\mathbfss{A}}_{\mathrm{half}}$, and let  $\mathbfit{C} \equiv \xoverline{\mathbfss{A}}_{\mathrm{half}}^{\mathrm{T}}\mathbfit{R}$, such that 

\begin{equation}
\mathbfss{B}\boldsymbol{\tau}_{\mathrm{half}} = \mathbfit{C}.
\label{eq:nonsquarematrixform_simple}
\end{equation}

We have therefore reduced our shadow imaging problem to the problem of solving a system of linear equations for the entries of the column vector $\boldsymbol{\tau}_{\mathrm{half}}$. These entries, re-shaped into the matrix $\tau$, correspond to the opacities of the pixels making up the top half of the grid, which define the image.

In the sections below, we elaborate upon the two steps of shadow imaging: first, selecting the grid parameters, and second, solving Equation~(\ref{eq:nonsquarematrixform_simple}) for the pixel opacities subject to our chosen physical constraints.

\subsection{Constraining the Grid Parameters}\label{subsec:gridParams}

In general, the auxiliary parameters $t_{\mathrm{ref}}$ and $v$ can be set to reasonable approximations of their ``true" values, and the pixel image will slightly shift or stretch, respectively, relative to the ``truth," without disturbance to its principal morphology. This means we can proceed by fixing these terms and optimizing the opacities $\boldsymbol{\tau}$ only. We may then, depending on the success of the solution $\boldsymbol{\tau}$, perform further iterations, varying the grid parameters each time, to reach an optimal grid with optimal auxiliary parameters. We discuss here some constraints of the grid parameters which allow us to estimate their values initially.

The first constraint we consider is that the number of
pixel elements should not exceed the number of data points obtained during
the transit event of interest. For regularly sampled data, such as that of \Kepler, we
may write the sampling constraint as

\begin{align}
N M \leq \frac{t_{\mathrm{event}}}{t_{\mathrm{cadence}}},
\label{eq:criterionA}
\end{align}

where $t_{\mathrm{event}}$ is the timescale of the event we wish to image and $t_{\mathrm{cadence}}$ is the cadence of the time series, i.e. the interval between successive observations.

The second constraint we consider is that a pixel should not be too small to detect individually. In other words, the transit depth of a single opaque pixel should not be smaller than the uncertainties on the flux measurements. In principle, smaller pixels could be resolved over repeated transit observations, but this approximation again aids in selecting a unique initial grid size from which to begin optimizing the grid opacities. 

Mathematically, we can express the precision constraint as:

\begin{equation}
    \frac{w^2}{\pi} \gtrsim \sigma
\end{equation}

where $\sigma$ is the typical photometric uncertainty. Combining Equation~(\ref{eq:pixelwidth}) with this
constraint gives

\begin{align}
N \lesssim \sqrt{\frac{4}{\pi \sigma}}.
\label{eq:criterionB}
\end{align}

For reference, using a 60\,ppm uncertainty, this yields $N \lesssim 146$. (In practice, we are usually limited to much smaller values of $N$ by the number of data points in the observed light curve.)

The third constraint we consider is the size of $M$. Our grid must be wide enough to create a total
duration sufficient to explain the event timescale, $t_{\mathrm{event}}$.
We require that $t_{\mathrm{exit}} - t_{\mathrm{enter}} \geq 
t_{\mathrm{event}}$, or

\begin{align}
\frac{2 + 2 (M/N)}{v} \geq t_{\mathrm{event}}.
\label{eq:criterionD}
\end{align}

%

Similarly, we consider that a single pixel needs to be able to traverse
the entire disk of the star within the event timescale. The actual
duration of a single pixel's transit will depend on the pixel's latitude $Y$, but to
simplify things, we consider an equatorial pixel of infinitesimal
size and use an approximate symbol for the inequality, to give

\begin{align}
v \gtrsim 2/t_{\mathrm{event}}.
\label{eq:criterionE}
\end{align}

Together, these expressions constrain the velocity to the range

\begin{align}
2/t_{\mathrm{event}} \lesssim v \leq  4/t_{\mathrm{event}}.
\end{align}

As a general strategy, then, we choose a grid velocity $v$ equal to $2/t_{\mathrm{event}}$, and $t_{\mathrm{ref}}$ to correspond to the minimum of the observed light curve. To choose $N$ and $M$, we recognize that, for a chosen $N$, we may solve for $M$ such that the grid continuously overlaps the star, by rearrangement of Equations~(\ref{eq:tBoundaries}). We can then adjust $N$ to accommodate the constraint that $NM$ be less than the number of observed data points. Once the grid dimensions have been chosen,
we re-execute the inversion for different velocities, until the fit
ceases to improve.

Because of the resolution constraint, we prefer the slowest grid velocity $v$ which returns a reasonable fit to the observed light curve, because this slow velocity corresponds to the highest image resolution $N$. This is, in a sense, an image prior which prefers narrow, slow images to their fast, stretch-degenerate counterparts.

\subsection{Solving for the Pixel Opacities}\label{subsec:pixelOpacities}

Once we have reasonable first estimates for $t_{\mathrm{ref}}$, $v$, $N$, and $M$, and have chosen a limb-darkening law to describe the stellar disk, we may use Equation~(\ref{eq:overlapareas}) to solve for $\bar{A}_{i,j}(t_k)$ for each grid pixel at each light curve time step. At this stage of shadow imaging,  it is helpful to think of the grid pixels as containers for as-yet-to-be-determined opacity: each transits the star in a definite way according to the grid parameters, so $\bar{A}_{i,j}(t_k)$ and hence $\xoverline{\mathbfss{A}}$ are well-defined, but its opacity is not yet known. 

For a chosen $t_{\mathrm{ref}}$, $v$, $N$, and $M$ and , we restrict our attention to the observed light curve data points that satisfy $t_{\mathrm{enter}} < t < t_{\mathrm{exit}}$. In other words, we consider only the points in time during which the grid partially overlaps the star, because the transiting grid could not influence points outside this range. 

To determine the opacities, we must solve Equation~(\ref{eq:nonsquarematrixform_simple}) for the entries of the opacity vector $\boldsymbol{\tau}$. Since this matrix equation is linear, in principle it can be directly, analytically solved. 

However, direct solution of Equation~(\ref{eq:nonsquarematrixform_simple}) cannot accommodate constraints on the pixel opacities. Namely, there is no way to restrict the entries of $\boldsymbol{\tau}$ to the physically meaningful range $[0,1]$ (case (1)), let alone to the binary values 0 or 1 (case (2)). Mathematically, introducing these constraints transforms the problem into a nonlinear optimization problem, which is not susceptible to solution by a linear matrix equation. We furthermore find that transforming the opacity variables through a logistic function, which maps the real numbers to the range $[0,1]$, results in numerical instabilities in our attempts to solve Equation~(\ref{eq:nonsquarematrixform_simple}) both directly and iteratively (e.g. with SART; see~\ref{subsubsec:SART} below).

Furthermore, we find that choosing grid parameters $t_{\mathrm{ref}}$, $v$, $N$, and $M$ that deviate even slightly from the true values leads to completely nonsensical recovered $\boldsymbol{\tau}$. Direct analytic solution is therefore not robust enough to apply to a light curve of unknown origin, where our initial guesses for the grid parameters are unlikely to be so accurate.

We therefore explore less exact, but significantly more robust, algorithmic approaches to solving for $\boldsymbol{\tau}$. Below, we discuss each of these algorithms in turn. The first two address case (1), where pixels may take on intermediate opacities, and the latter three address the more restrictive case (2), where pixels are constrained to be binary-valued.

\begin{figure*}[t!]
\begin{center}
\includegraphics[width=1.02\textwidth]{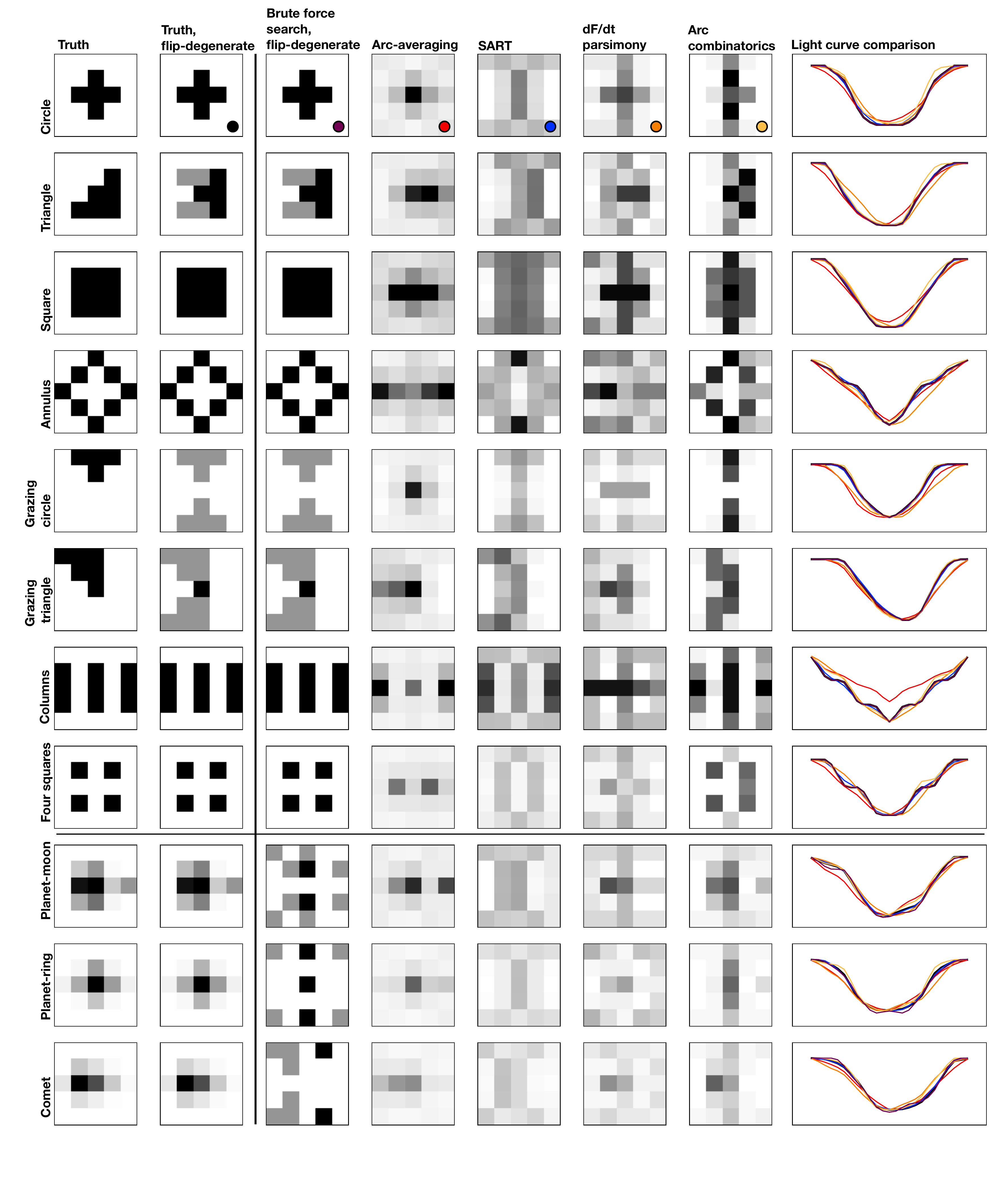}
\caption{The performance of several light-curve inversion algorithms on eleven known 5 by 5-pixel test grids. The leftmost two columns represent the true input grid; the subsequent columns represent the grid recovered by each inversion algorithm given only the (noiseless) true light curve as input. The eight test grids above the horizontal line are pure binary grids (i.e., pixel opacities are either 0 or 1); the three below have intermediate, semi-opaque pixels. Each algorithm was initialized with correct grid parameters $N$, $M$, $t_{\mathrm{ref}}$, and $v$, and the light curves were generated with a uniform limb darkening law. The brute-force search algorithm performs the best, i.e. returns the light curve with lowest RMS error compared to the true image's light curve, in every pure binary test case, but SART performs best on the semi-opaque test cases.}
\label{fig:5sq}
\end{center}
\end{figure*}

\begin{figure*}[t!]
\begin{center}
\includegraphics[width=1.02\textwidth]{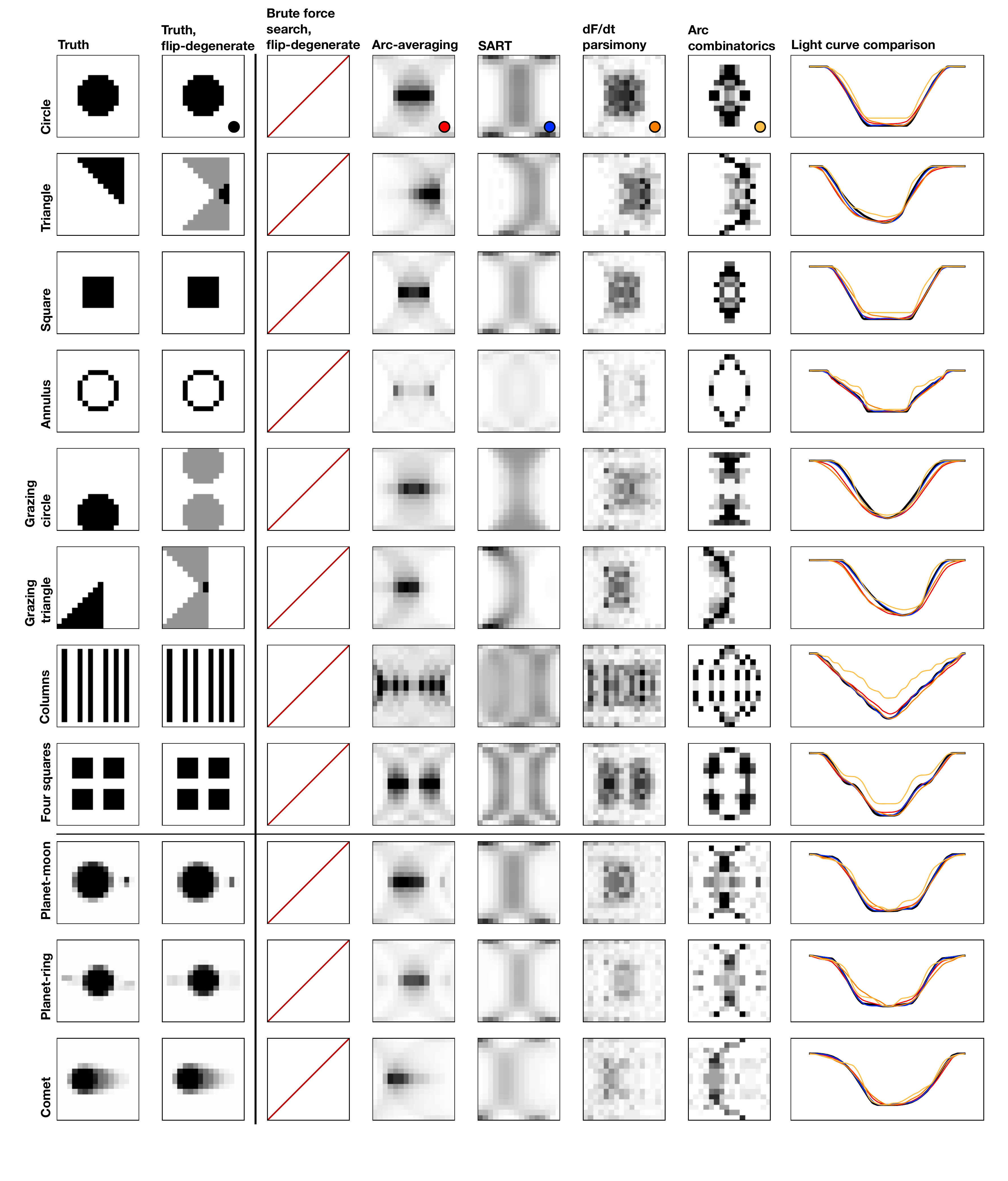}
\caption{The performance of several light-curve inversion algorithms on eleven known 16 by 16-pixel test grids, which are too large to allow for a brute-force permutation search. The leftmost two columns represent the true input grid; the subsequent columns represent the grid recovered by each inversion algorithm given only the (noiseless) true light curve as input. The eight test grids above the horizontal line are pure binary grids (i.e., pixel opacities are either 0 or 1); the three below have intermediate, semi-opaque pixels. Each algorithm was initialized with correct grid parameters $N$, $M$, $t_{\mathrm{ref}}$, and $v$, and the light curves were generated with a uniform limb darkening law. SART performs the best, i.e. returns the light curve with lowest RMS error compared to the true image's light curve, in every test case; arc-averaging is second-best in every case except the offset circle, for which arc combinatorics does better.}
\label{fig:16sq}
\end{center}
\end{figure*}

In Figures~\ref{fig:5sq} and~\ref{fig:16sq}, we compare their performances in recovering a number of known test grids from noiseless light curves. In these recovery tests, the parameters $N$, $M$, $v$, and $t_{\mathrm{ref}}$ were assumed to be known. Eight of the test grids are binary-valued, and three (the low-resolution planet-moon, planet-ring, and comet) include intermediate-opacity pixels.

In Figures~\ref{fig:5sq} and~\ref{fig:16sq}, we have chosen to generate our test light curves with a uniformly bright star, i.e., without limb darkening. We make this choice because non-limb-darkened light curves are sharper and less rounded than limb-darkened light curves, and the inversions result in correspondingly sharper image grids, among which the differences between the images generated by our four recovery algorithms stand out most clearly.

We find that introducing realistic limb darkening results in very similar recovered images to those shown in Figures~\ref{fig:5sq} and~\ref{fig:16sq}, with two notable qualitative differences: first, for the limb-darkened case, opacity tends to be pushed farther out towards the top and bottom edges of the recovered image. This effect is most obvious in the arc-combinatoric images. Second, the recovered images appear blurrier, which makes intuitive sense given the more rounded features of a limb-darkened transit event compared to a non-limb-darkened transit.

\subsubsection{Arc-Averaging}\label{subsubsec:arcAveraging}

The first algorithmic approach we explore relies on the time derivative of Equation~(\ref{eq:matrixform}). At each time step $dt$, the overlap state of the grid changes; we can express the change in overlap area as the matrix $d\xoverline{\mathbfss{A}}/dt$, calculated at each time step. Most of the entries of this matrix will be equal to 0, because only the pixels overlapping the stellar limb at that time step will have nonzero change in overlap area.

Meanwhile, at each $dt$, we can calculate the net change in the observed light curve, $d\mathbfit{R}/dt$. Two effects can contribute to nonzero $d\mathbfit{R}/dt$ at a particular time step: (i) one or more pixels with nonzero opacity overlapping the stellar limb at that time step, and (ii) in the case of non-uniform limb darkening, one or more pixels with nonzero opacity overlapping any part of the star. For the low-resolution grid inversions possible given the time resolution of currently available transit data (see e.g.~\ref{subsec:trappist1} and~\ref{subsec:boyajian}), effect (i) is much larger than effect (ii). Additionally, the stellar intensity profile changes most steeply near the limb, so effect (ii) is most prominent for limb pixels anyway.

Therefore, for the ``arc-averaged" algorithm, we take the naive approach of calculating the average $d\mathbfit{R}/dt$ per pixel which overlaps the limb at that time step. Then, we endow each limb pixel with that average opacity, weighted by $\frac{1}{\sin{\theta_{\mathrm{pixel}}}} = \frac{R_*}{b_{\mathrm{pixel}}} = \frac{1}{b_{\mathrm{pixel}}}$ to mitigate the effects of the arc degeneracy.

We do the above arc-averaging independently for each time step $dt$, then average the results over all time steps to compute the final grid. Finally, we re-normalize the pixel grid to match the transit depth of the observed light curve. (Renormalization is necessary because the arc-averaging algorithm only exploits information from the derivative of the light curve, not from the light curve itself.)

Arc-averaged pixel solutions, because they exploit the arc degeneracy, exhibit semicircular arc-like features. They are also horizontally symmetrical as a result of the flip degeneracy. Overall, they are smoother and more dispersed than their true pixel grid counterparts, with smoother light curves, because the averaging step precludes sharp, isolated islands of opacity. The impact parameter weighting causes opacity to be concentrated at the midplane of the grid.

As shown in Figures~\ref{fig:5sq} and~\ref{fig:16sq}, the light curves of arc-averaged solutions match observed light curves well, particularly for large, centrally-concentrated test shapes. The worst matches are for grazing shapes (see e.g. the 16 by 16-pixel grazing circle), because the $\frac{1}{b_{\mathrm{pixel}}}$ weighting pushes opacity strongly toward the grid midplane and away from the top and bottom of the grid. The arc-averaged light curves also tend to be more rounded than the observed light curve, meaning that the arc-averaging approach struggles to reproduce sharp light curve features. This is sensible because, by design, it produces solutions where opacity is distributed continuously along overlapping arcs rather than confined to discrete islands.

We also note that, because arc-averaging can easily accommodate pixel opacities between 0 and 1, it can be applied to semi-opaque pixel grids, like the low-resolution planet-moon, planet-ring, and comet test grids.

\subsubsection{Simultaneous Algebraic Reconstruction Technique (SART)}\label{subsubsec:SART}

The next algorithm we test is called the Simultaneous Algebraic Reconstruction Technique, or SART \citep{andersen84}. SART was originally developed for medical computed tomographic imaging. Specifically, SART reconstructs a 2D image from the projections of X-rays through the body---this is directly analogous to our shadow imaging problem, where the ``projections" of the pixelated image against the stellar disk are the individual data points in the light curve. 

SART operates iteratively upon an initial guess for the opacity vector $\boldsymbol{\tau}_{\mathrm{half}}$, which encodes the opacities of the pixels of the top half of the image grid. Beginning from this initial guess, it computes subsequent corrective updates to the individual entries of $\boldsymbol{\tau}_{\mathrm{half}}$. 

The $(q+1)^{\mathrm{th}}$ iteration of $\tau_l$, the opacity of pixel $l$, is given by



\begin{equation}
    \tau_l^{q+1} = \tau_l^q + \frac{\sum\limits_{k=1}^{L_{\mathrm{half}}} \left(B_{kl} \frac{C_k - \sum\limits_{\lambda=1}^{L_{\mathrm{half}}} (B_{k\lambda}\,\tau_{\lambda}^q) }{\sum\limits_{\lambda=1}^{L_{\mathrm{half}}} B_{k\lambda}} \right)}{\sum\limits_{k=1}^{L_{\mathrm{half}}} B_{kl}},
\label{eq:SART}
\end{equation}

The scalar $\tau_l^q$ is the $l^{\mathrm{th}}$ entry of $\boldsymbol{\tau}_{\mathrm{half}}$ at iteration $q$, representing the opacity of pixel $l$; the scalar $B_{kl}$ is $k^{\mathrm{th}}$-row, $l^{\mathrm{th}}$-column entry of $\mathbfss{B}$; and the scalar $C_k$ is the $k^{\mathrm{th}}$ entry of $\mathbfit{C}$. $\mathbfss{B}$ and $\mathbfit{C}$ are defined in Equation~(\ref{eq:nonsquarematrixform_simple}).

For intuition, the update term in Equation~(\ref{eq:SART}) is equal to the average correction to pixel opacity $\tau_l$ over all rows and all columns of $\mathbfss{B}$. (Hence, the sum in the denominator is over all rows of column vector $\mathbfit{B}_l$, and the sum in the numerator term's denominator is over all columns of row vector $\mathbfit{B}_k$.) The numerator, specifically, is the average value over all pixels in the grid of a sort of ``residual" between the observed light curve and the model. This residual is equal to $C_k$ minus the scalar projection of $\boldsymbol{\tau}^q_{\mathrm{half}}$ along $\mathbfit{B}_k$. In effect, these two averages allow for a correction to the opacities which is averaged over all time steps of the light curve and all pixels in the grid.

By running the SART algorithm for a large number of iterations (usually, $\sim 10^4$ for a 16 by 16 pixel grid), we achieve good convergence to the observed light curve for a number of test cases. The RMS error between the light curve of the input image and the light curve of the SART solution declines monotonically over the SART iterations, indicating that SART achieves a progressively better fit to the light curve as it proceeds.

We find that starting from an initial guess of all $\tau_l = 0.5$ works well, because the step-by-step updates to $\boldsymbol{\tau}$ are generally small, so the algorithm does not wander far into unphysical parameter space (i.e., $\tau_l < 0$ or $\tau_l > 1$). In the event that the resulting SART solution does have slightly unphysical opacities, we redistribute the excess positive or negative opacity uniformly among the pixels whose centers fall within a distance of $w/2$ of the arc pair that intersects at the unphysical pixel. This redistribution renders the SART solution fully physical without drastically changing its light curve. Because SART exploits information in the light curve, not just its derivative, it is not necessary to re-normalize the SART solution pixel opacities.

SART solutions exhibit horizontal symmetry as a result of the flip degeneracy, and semicircular arc-like features as a result of the arc degeneracy. Like the arc-averaging algorithm, SART tends to smear out sharp features in the true input image along arcs, resulting in pixel grid solutions which are smoother, with more dispersed opacity than the true image. (SART solutions are even smoother than the corresponding arc-averaged solutions.) As as a result, SART fails, for example, to match the sharply flat-bottomed transits of the 16 by 16-pixel circle and square test grid light curves (Figure~\ref{fig:16sq}), producing slightly rounded light curve shapes instead. On the other hand, because SART allows the pixel opacities to take any continuous value between 0 and 1, it can accurately reproduce the light curves of non-binary test grids, like the planet-moon, planet-ring, and comet.

\subsubsection{Brute Force Search}\label{subsubsec:bruteForce}

The next three algorithms we explore attempt to invert light curves subject to the constraint of binary pixel opacities: in other words, we attempt to recover grids with pixel opacities of 0 (completely transparent) or 1 (completely opaque). We begin with the simplest, a brute-force search of every possible arrangement of binary pixel opacities.

As discussed in~\ref{subsec:flipDegeneracy}, by the flip degeneracy, a grid of $N$ by $M$ opaque and transparent pixels can generate $\mathcal{O}[3^{NM/2}]$ unique light curves. Correspondingly, one would have to evaluate $\mathcal{O}[3^{NM/2}]$ permutations of transparent and opaque pixels to find the grid that matches a given light curve best. The largest square grid for which such a full search is feasible is 5 by 5 pixels, which has 1.9 million associated pixel arrangements with unique light curves (for comparison, a 6 by 6 grid has $\sim 390$ million).

In Figure~\ref{fig:5sq}, we illustrate the results of a brute force full-grid search for noiseless test light curves generated by number of 5 by 5 known input grids. The brute force algorithm returns the pixel arrangement which, when transiting the star, generates a light curve with the lowest RMS error compared to the truth. 

Unsurprisingly, when the input grid is truly binary, i.e. made up of completely opaque and completely transparent pixels, the full search converges to the best possible solution every time. However, when the input grid includes semi-opaque pixels, as in the low-resolution planet-moon, planet-ring, and comet test cases, the brute force search struggles; the lowest-RMS solution does not necessarily bear any resemblance to the input grid, even though its light curve matches the true light curve well. This is a testament to the complex and multi-modal likelihood landscape of the pixel opacities, and also an illustration of why conventional nonlinear optimization methods cannot solve the light curve inversion problem. (We note here that we also investigated both a genetic algorithm and a downhill simplex algorithm \citep{neldermead65}, without success---both methods tended to reach local optima and stall, and as illustrated here, locally optimal grids are not necessarily morphologically similar to the true grid.)

Brute-force search solutions are not presented in Figure~\ref{fig:16sq}, because these grids are far too large to be exhaustively permuted.

\subsubsection{Parsimonious Opacity Assignment}\label{subsubsec:parsimony}

The next two algorithms we test rely, like arc-averaging, on the time derivative of Equation~(\ref{eq:matrixform}). However, instead of averaging the ingress or egress opacity over all of the limb pixels at each time step, we attempt to parcel it out in units of 0.5 opacity (to accommodate the flip degeneracy). We note that consequently, these two algorithms do not work well for inverting shallow transits observed with low time sampling (i.e., few light curve data points), because in such cases, the grid will be low-resolution, and the transit depth of a single pixel's worth of opacity can be greater than the observed transit depth. There will then be no good match to the light curve.

First, we explore the ``parsimonious" approach, which assigns opacity to as few pixels as possible in order to accommodate the change in the light curve. This algorithm is motivated by compactness--is it possible to match the light curve with as few ``on" pixels as possible?

The parsimonious approach assigns opacity first to the pixel with the largest change in overlap area $d\xoverline{\mathbfss{A}}/dt$, then steps through successively ``less influential" pixels until the entire change in the light curve has been accounted for. As with the arc-averaging approach, it is necessary to average the results over all time steps $dt$, then renormalize the resulting pixel grid to match the observed transit depth; the pixel grid solutions presented in Figures~\ref{fig:5sq} and~\ref{fig:16sq} therefore have some pixel opacities between 0 and 1.

In practice, this algorithm generates pixel grids which are strongly concentrated at the stellar midplane, because these middle pixels undergo the greatest change in overlap area at fixed $dt$ during their ingress and egress. Correspondingly, it fails to reproduce high-impact-parameter features in the input grids, and is especially poor at matching the light curves of grazing shapes, like the grazing circle and grazing triangle (Figure~\ref{fig:16sq}). Overall, it is the least successful of the four algorithms. 

\subsubsection{Arc Combinatorics}\label{subsubsec:arcCombinatorics}

Finally, we consider an algorithm which attempts to assign units of 0.5 opacity to the best \textit{combination} of limb pixels at every time step in order to match the observed light curve. At every $dt$, the algorithm calculates the number of ``spaces" on the stellar limb, $s$, that could accommodate a unit of 0.5 opacity. This is equal to twice the number of limb pixels of the appropriate ``sign:" if the light curve is decreasing at $dt$, we need only consider the limb pixels which are undergoing ingress, and vice versa.

Next, it calculates the number of 0.5-opacity units $n$ that need to be accommodated. This is equal to the change in the light curve, $d\mathbfit{R}/dt$, divided by the mean overlap area of the limb pixels at that time step, multiplied by 2 (because we wish to distribute opacity in units of 0.5, not 1).

The number of ways to arrange $n$ opacity units over $s$ spaces is then ${s}\choose{n}$. The algorithm explores each combination and chooses the one which matches the vector $d\mathbfit{R}/dt$ best. Finally, as with arc-averaging and the parsimonious approach, the resulting grid is averaged over all time steps and renormalized to match the observed transit depth (so once again, the pixel grid solutions presented in Figures~\ref{fig:5sq} and~\ref{fig:16sq} therefore have some pixel opacities between 0 and 1).

The arc combinatorics approach is able to match certain vertically-sharp features in the input images, such as the 16 by 16-pixel annulus and column test cases (Figure~\ref{fig:16sq}). It can also accommodate narrow features at high impact parameter; to see this, compare the parsimonious and arc combinatorics solutions to the 16 by 16-pixel four-squares test case. 

Because of the arc degeneracy, however, the arc combinatorics algorithm tends to prefer solutions where opacity is pushed too far toward the top and bottom edges of the grid (e.g. the 16 by 16-pixel circle and square test grids, Figure~\ref{fig:16sq}). (This is the opposite problem of the parsimonious algorithm.) It also struggles to capture the nuances of semi-opaque test grids, like the planet-moon, planet-ring, and comet. Finally, we note that the computational cost of this algorithm scales poorly with increasing grid resolution (i.e., increasing $s$), because the algorithm needs to evaluate ${s}\choose{n}$ opacity arrangement possibilities.


\section{Real Data}
\label{sec:realData}

In this section, we discuss the performance of shadow imaging on two real test cases: first, the light curve of the triple transit of TRAPPIST-1c, e, and f \citep{gillon17}, and second, two unexplained transit-like events observed in KIC 8462852, or Boyajian's Star \citep{boyajian16}. 

\subsection{TRAPPIST-1c,e,f triple transit} \label{subsec:trappist1}

\begin{figure*}[t!]
\begin{center}
\includegraphics[width=0.8\textwidth]{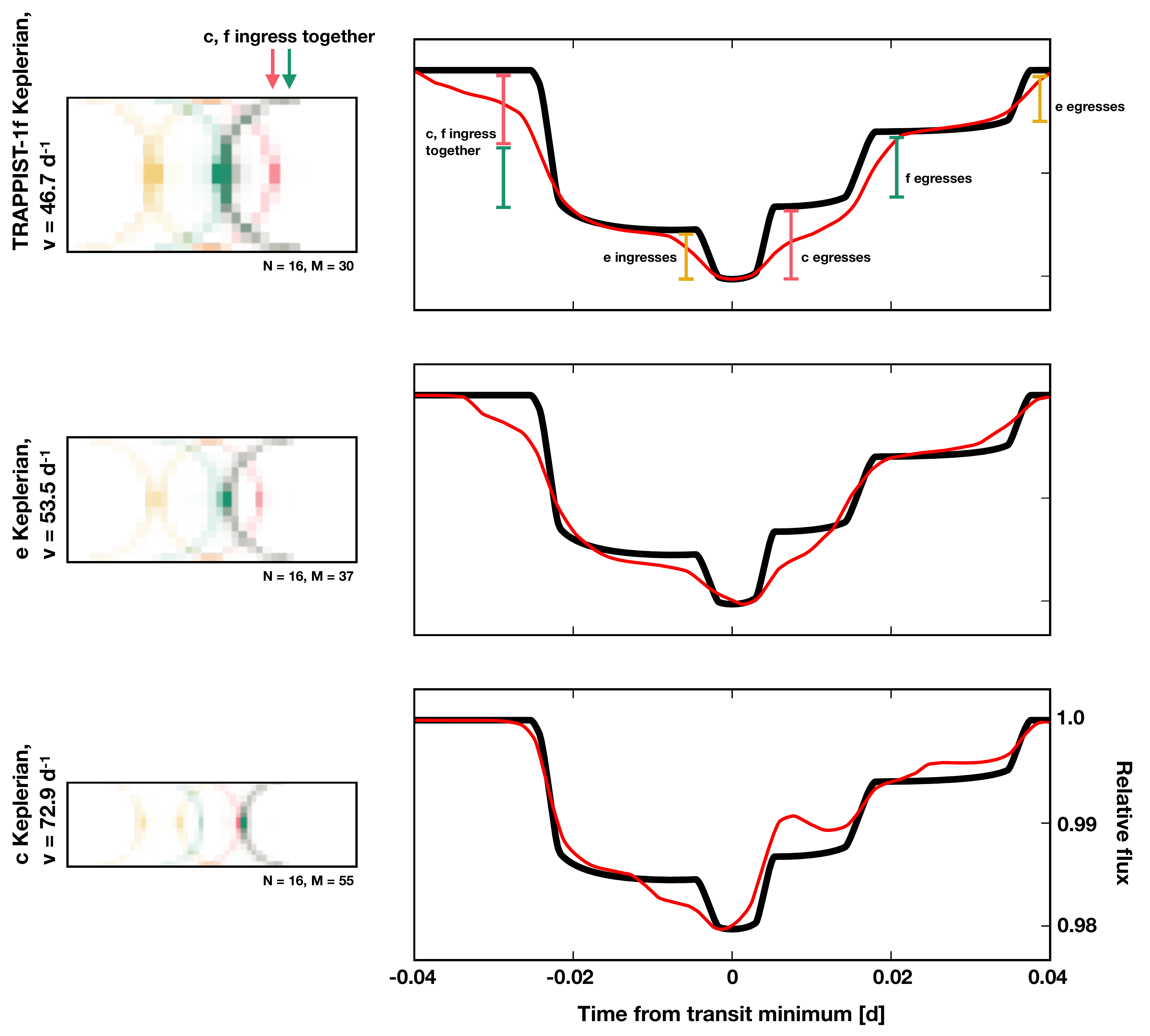}
\caption{Three inversions of a BATMAN-modeled, high-resolution TRAPPIST-1c,e,f, triple transit light curve, conducted with the arc-averaging algorithm, with grid $v$ equal to the Keplerian velocity of planets c (bottom), e (middle), and f (top). These images transit the star moving left to right, so the features at the right-hand side of the image influence the light curve first. The BATMAN model light curve (black) and arc-averaged shadow image light curve (blue) are compared in the right-hand panels. We have added color to the shadow images to indicate the positions of planets c (pink), e (yellow), and f (green). (Note that c and f ingress together, so their ingress arc is green + pink = gray.)}
\label{fig:trappist1model}
\end{center}
\end{figure*}

\begin{figure*}[t!]
\begin{center}
\includegraphics[width=0.9\textwidth]{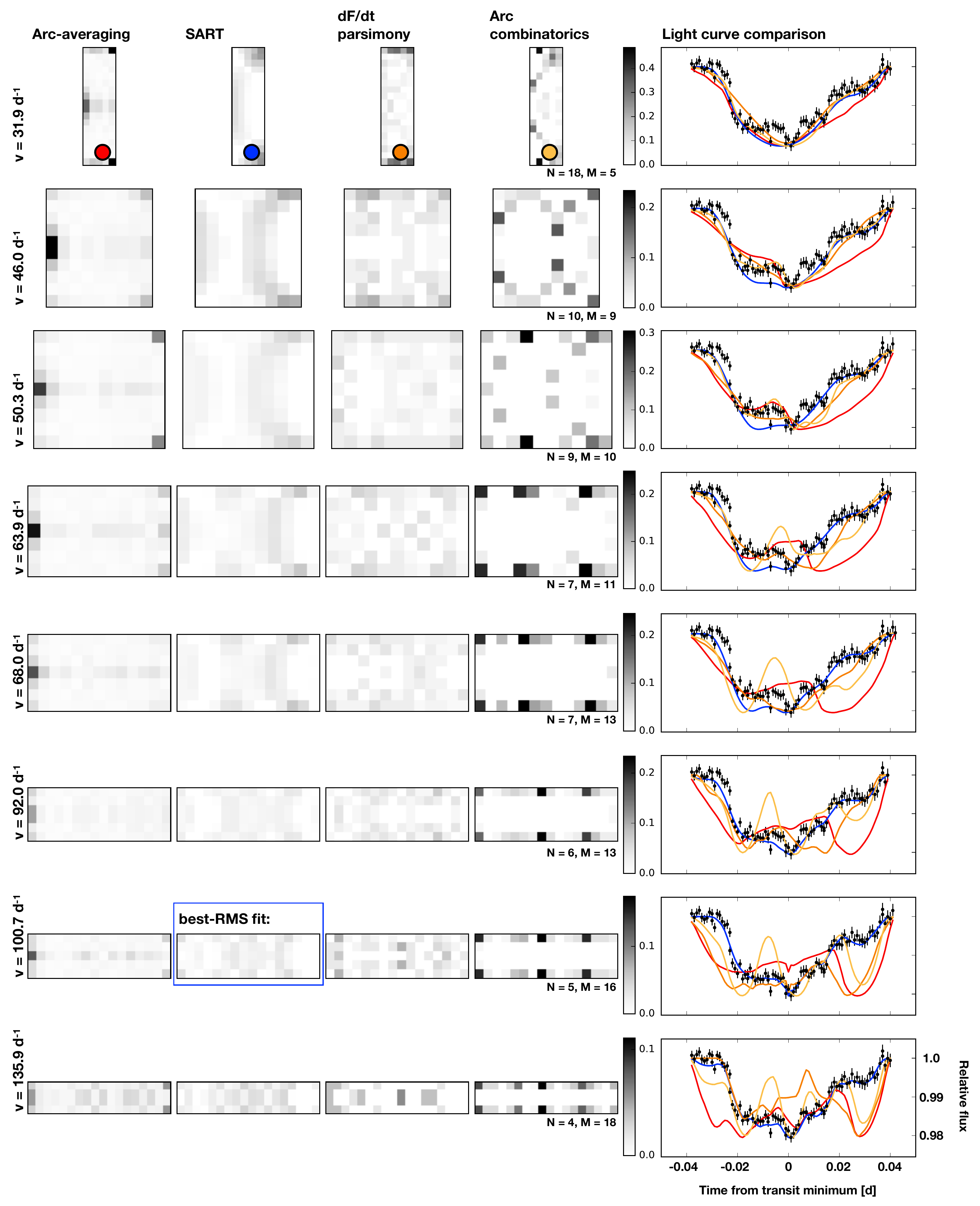}
\caption{The performance of several light-curve inversion algorithms on the observed TRAPPIST-1c,e,f triple transit light curve. The test velocities and corresponding grid resolutions were chosen according to the guidelines set out in~\ref{subsec:gridParams}. The shadow image whose light curve has the lowest RMS error compared to the observed light curve is the SART inversion at $v = 100.7 \mathrm{d}^{-1}$, marked by the blue box. Arc combinatorics performs best, by RMS, at the two slowest tested velocities, but SART performs best at all the others.}
\label{fig:trappist1obs}
\end{center}
\end{figure*}

We begin with the TRAPPIST-1c,e,f triple transit, for which the expected shadow image is known. We hope to recover an image of three transiting planets, analogous to the diagram presented in \cite{gillon17} Extended Data Figure 1. 

In attempting to invert this light curve, we have useful prior information beyond the expected image. First, because of the repeated transit observations and N-body dynamical simulations presented in \citep{gillon17}, the periods and eccentricities, respectively, of planets c, e, and f are well-constrained. This enables us to calculate the Keplerian orbital velocities of c, e, and f, which we can use as $v$ of our transiting grid. (We note that since these three orbital velocities are different, the pixel image we are attempting to recover changes during the transit, so we will only be able to recover an approximate image for any single choice of $v$.)

Second, because the physical behavior of this system is so well-understood and the other properties of these planets ($R_p/R_*$, $b$, $a/R_*$) are so well-constrained by transit modeling, we can generate an extremely finely time-sampled model light curve, based on a BATMAN model \citep{kreidberg15}, of this triple transit, which matches the observed light curve. We can use this high-resolution light curve to test the effects of grid resolution on the success of shadow imaging: when the light curve is finely sampled, we can recover a much higher-resolution grid than when the light curve is sparsely sampled. Finally, we can adopt the same approach to determining the quadratic limb-darkening coefficients for TRAPPIST-1 as \cite{gillon17} did in their analysis, interpolating for TRAPPIST-1's stellar properties from the tables of \cite{claret12}.

In Figure~\ref{fig:trappist1model}, we present three inversions of the BATMAN-modeled high-resolution TRAPPIST-1c,e,f, triple transit light curve, conducted with grid $v$ equal to the Keplerian velocity of planets c, e, and f, respectively. We choose $N$ = 16 because it is a high-enough resolution that pixel width $w \lessapprox R_p/R_*$ for planet e, the smallest planet ($(R_p/R_*)^2 = 0.52$, according to the transit modeling of \citealt{gillon17}). We show the results of the arc-averaging algorithm here, because it produces the cleanest and most interpretable shadow images, although results from the other three algorithms are qualitatively similar.

In the shadow images, which transit the star moving left-to-right (i.e., the pixels at the right-hand edge of the image transit first), clear ingress and egress arcs for each planet are visible, in the expected order: first, planets c and f ingress together; then, e ingresses; c egresses; f egresses; and e egresses. 

The three planets move at three different velocities to produce the light curve, but the grid moves as a unit, so none of the three inversions perfectly matches the light curve model. When the velocity is correct for a particular planet, that planet's image is a pair of arcs whose points of intersection fall at the planet's impact parameter as measured by \cite{gillon17}, demonstrating that shadow imaging of that planet is successful within the constraints of the arc and flip degeneracies. 

When the grid $v$ is slower than the planet's velocity, the planet's ingress and egress arcs are spaced too closely together; this effect is most visible for planet c in the top panel, where the grid moved at planet f's velocity. In the light curve, the overlapping arcs manifest themselves in a too-early dip, caused by c's \textit{egress} arc entering too quickly, and in a too-deep transit depth between the egresses of c and f, caused by c's \textit{ingress} arc remaining in front of the star for too long.

Conversely, when the grid velocity is faster than the planet's, the planet's arcs are too widely separated; this effect is most visible for planet f in the bottom panel. This time, the light curve is too shallow between the egresses of planets c and f, because f's \textit{ingress} arc egresses too soon.

We next investigate what happens if we attempt to invert the observed \cite{gillon17} light curve of this triple transit, which is noisy and much more coarsely time-sampled, rather than a high-resolution BATMAN model light curve. Additionally, we ask what happens if we attempt to recover a shadow image without knowing the true velocity of the transiting object: what happens if we use the guidelines presented in~\ref{subsec:gridParams} instead?

We invert the  observed TRAPPIST-1 triple transit light curve at a range of velocities: the slowest is $31.9\ \mathrm{d}^{-1}$, corresponding to $2$ divided by the entire triple-transit event duration (in accordance with the guidelines presented in~\ref{subsec:gridParams}), and the fastest is $135.9\ \mathrm{d}^{-1}$, corresponding to $4$ divided by the duration of planet c's transit by itself. At each velocity, we choose the maximum grid resolution $N$ that, when combined with $v$ to solve for $M$, allows the transiting grid to partially overlap the star at all time steps of the light curve, while still maintaining $NM$ less than the number of observed data points. Accordingly, the resolution $N$ decreases as $v$ increases, because $M$ increases with $v$ to maintain full light curve coverage.

In Figure~\ref{fig:trappist1obs}, we present the results of these inversions. There are a number of interesting features about these results. We note, first of all, that SART is consistently the most successful inversion algorithm---this is true across the range of tested grid velocities. Furthermore, the SART shadow image consistently resembles the expected shadow image illustrated in Figure~\ref{fig:trappist1model}, even at low image resolutions. Arc combinatorics is somewhat successful at matching the observed light curve at the slowest tested velocity (corresponding to the highest grid resolution), but fails otherwise.

The other algorithms fail consistently across the range of tested velocities. For arc parsimony and arc combinatorics, this results because these algorithms assign binary opacities (0 or 1) to individual pixels, rather than assigning continuous opacities. (We note that the shadow images presented in Figure~\ref{fig:trappist1obs} do not have binary opacities because the final step of both the arc parsimony and arc combinatorics algorithms is to average the binary shadow images produced at each time step $dt$ and re-normalize the average to match the observed transit depth.) 

When the pixel resolution of the grid is too low, a single pixel's transit depth can exceed the transit depth of a shallow event like the TRAPPIST-1c,e,f triple transit (maximum transit depth $\sim 2\%$). As a result, the smallest unit of opacity that the arc parsimony or arc combinatorics algorithms can assign is too deep, and these algorithms cannot reproduce the observed light curve. Instead, they tend to assign opacity to pixels along the top and bottom of the image grid, which have the smallest impact on the light curve. This is especially visible in the high-$v$ arc combinatorics panels in Figure~\ref{fig:trappist1obs}.

Meanwhile, the arc averaging algorithm also fails to match the observed light curve, regardless of velocity. This is because the arc-averaging algorithm, unlike SART, is not robust to noise in the light curve; noise is tantamount to light-curve fluctuations at much higher frequency than can be accommodated by the grid velocity. While SART is able to average out high-frequency noise over many corrective iterations, arc averaging calculates only one arc arrangement per time step $dt$; if these arrangements are wildly different for neighboring time steps, as they will be for noisy light curves, arc averaging fails.

From these investigations, we conclude that SART is the most robust light curve imaging algorithm. In particular, light curves with large measurement uncertainties and/or shallow transit depths should only be inverted with SART.

\subsection{Boyajian's Star}\label{subsec:boyajian}

\begin{figure*}[t!]
\begin{center}
\includegraphics[width=0.95\textwidth]{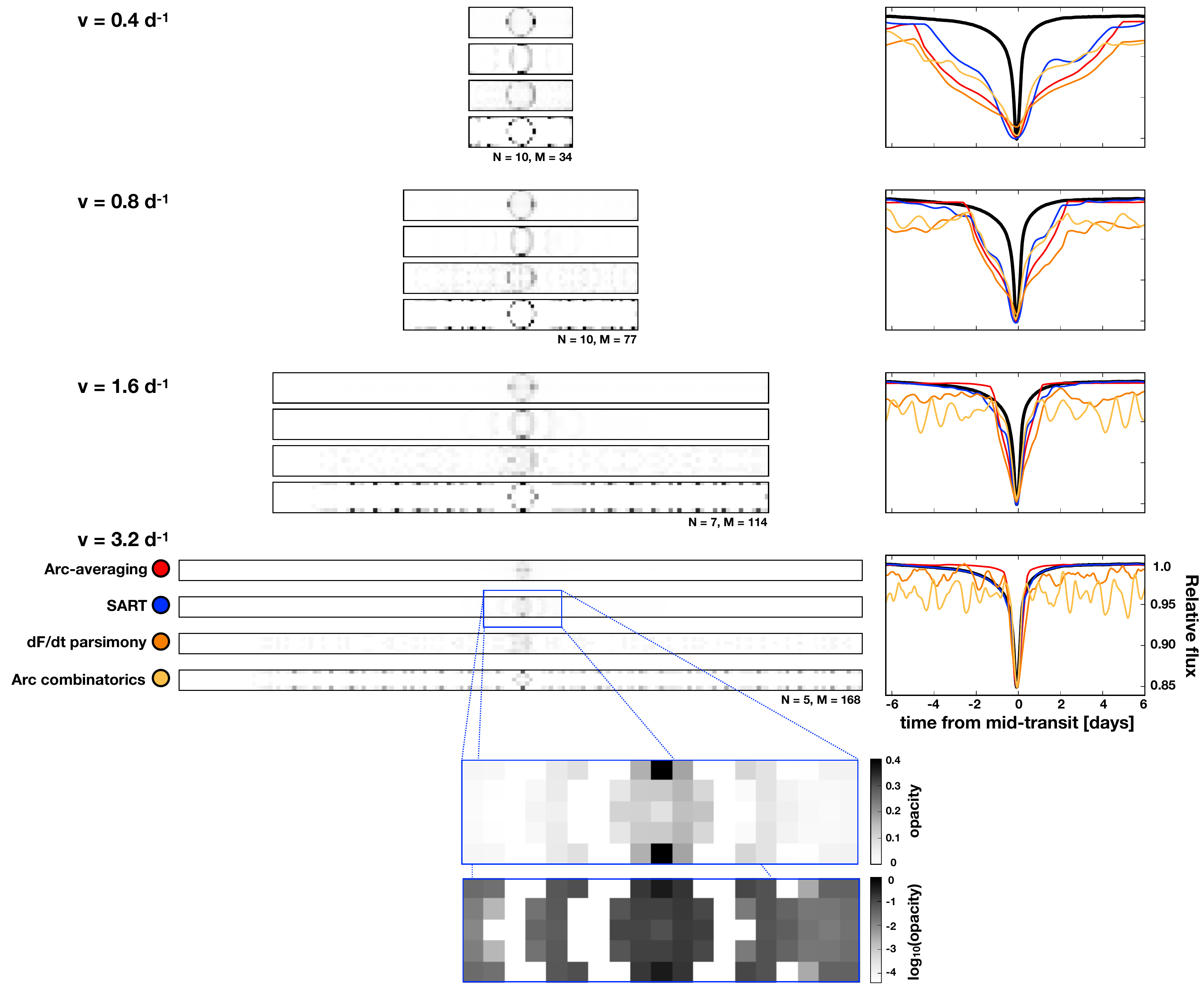}
\caption{The performance of several light-curve inversion algorithms on Dip 5 of Boyajian's Star. Inset: a zoomed-in view of the central SART shadow image, with both linear and logarithmic color scaling to represent opacity. SART performs best, by RMS, at all four choices of $v$.}
\label{fig:boyajian_dip5}
\end{center}
\end{figure*}

\begin{figure*}[t!]
\begin{center}
\includegraphics[width=0.7\textwidth]{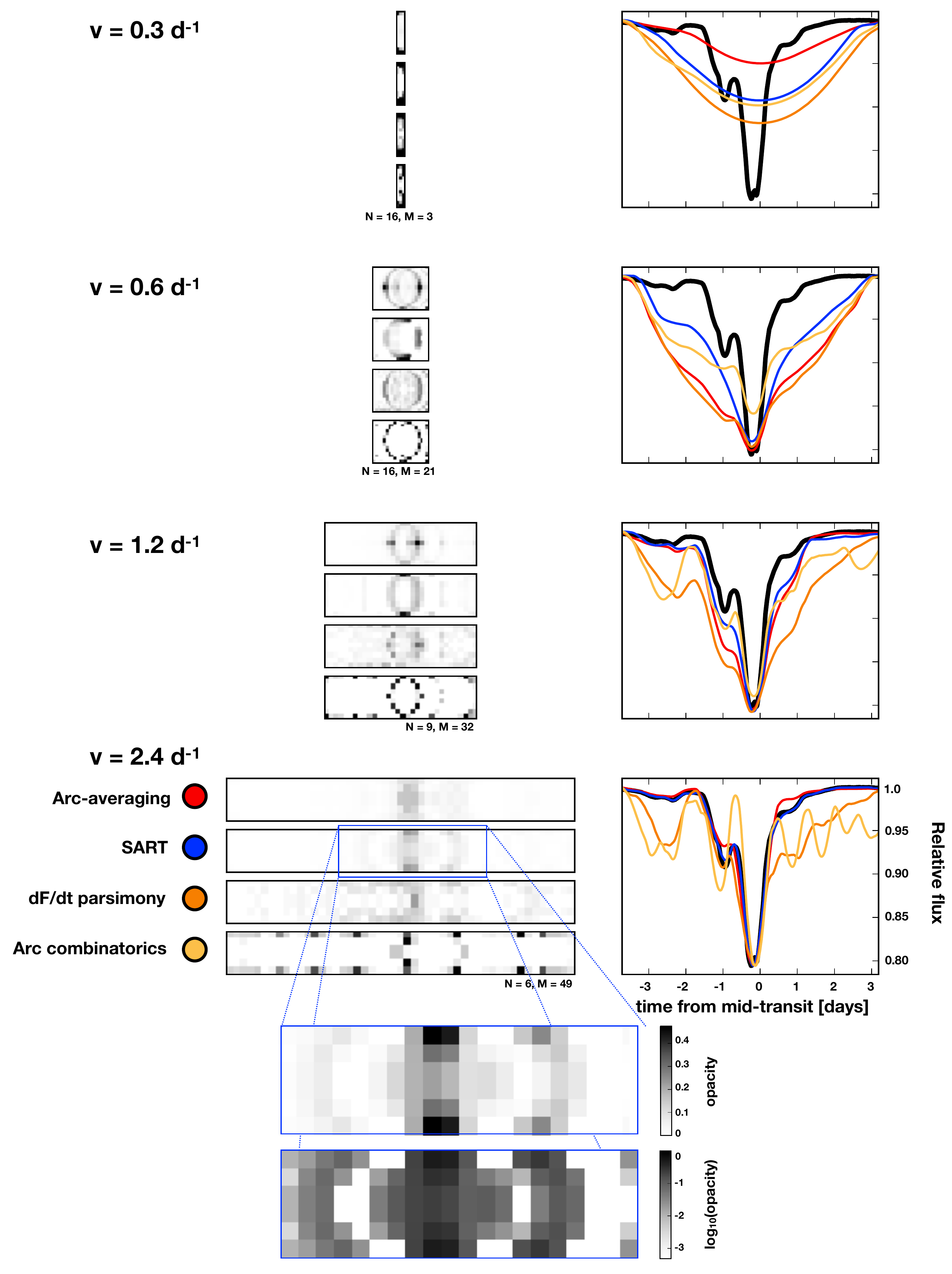}
\caption{The performance of several light-curve inversion algorithms on Dip 8 of Boyajian's Star. Inset: a zoomed-in view of the central SART shadow image, with both linear and logarithmic color scaling to represent opacity. SART performs best, by RMS, at all four choices of $v$.}
\label{fig:boyajian_dip8}
\end{center}
\end{figure*}

Next, we proceed to a light curve with an unknown generative shadow image: that of KIC 8462852, Boyajian's Star \citep{boyajian16}. This star exhibits aperiodic, deep transit events of unknown origin; hypotheses to explain these events include a family of transiting comets \citep{boyajian16,bodman16}; circumstellar rings \citep{katz17}; an intervening occulter not orbiting Boyajian's Star directly, such as structure in the interstellar medium or an object with a dusty disk \citep{wright16}; or circumstellar debris following the star's earlier engulfment of a planet \citep{metzger17}; or alien megastructures orbiting the star \citep{wright16megastructures}.

We focus on the two deepest dimming events observed in the Boyajian's Star light curve during the \textit{Kepler} mission, which \cite{boyajian16} label Dip 5 and Dip 8, respectively. Since these events are aperiodic, the appropriate grid velocity $v$ is not obvious; furthermore, in both dips, the light curve smoothly tapers to a sharp point, so the ``beginning" and "ending" points of the event are not obvious. Correspondingly, we start from a velocity $v = 2/t_{\mathrm{event,\ max}}$, where $t_{\mathrm{event,\ max}}$ is a wide window around the deepest part of the transit, outside of which the flux of the star has essentially returned to 1 again. (These time ranges are plotted in Figures~\ref{fig:boyajian_dip5} and~\ref{fig:boyajian_dip8}). We then test several other velocities doubled from this starting point. We interpolate quadratic limb-darkening coefficients for Boyajian's Star from \cite{sing10}.

The inverted images for Dips 5 and 8 are presented in Figures~\ref{fig:boyajian_dip5} and~\ref{fig:boyajian_dip8}, respectively. There are a number of interesting features in these images. 

First, there is a circular ring-like feature, of the same radius as the star, that appears generally in images from all four algorithms when the grid velocity $v$ is too slow. (See Dip 5, $v \leq 1.6$ days, and Dip 8, $v \leq 1.2$ days, for examples.) This happens because, when the grid velocity is too slow, the grid struggles to produce narrow features in the light curve; the rate of change of the state of the grid overlap is simply too slow. Under this constraint, the circular ring is the grid pattern that matches a narrow light curve feature best, in the sense that it generates the narrowest possible V-shaped light curve feature. 

For intuition, consider a copy of the circular ring with the addition of some opaque interior pixels: at some time steps, these interior pixels will be entirely contained within the stellar disk, and their effect on the light curve during these time steps will be constant. In other words, their transit will be flat-bottomed. This is not the case for the ring, whose overlap state changes at every time step of the transit; the ring is the ``opposite" of a flat-bottomed semicircular arc pair in this sense. 

Velocities which produce a ringed shadow image are therefore too slow. This is also obvious from the light curve of the shadow image, which is wider than the observed transit event. 

When $v$ is fast enough, we find that all four algorithms produce qualitatively similar shadow images for both Dip 5 and Dip 8, and furthermore that the shadow images of the two dips are similar to each other. As in the case of the TRAPPIST-1 triple transit, the arc parsimony and arc combinatorics algorithms generate ``noisy" shadow images where the light curve is shallow, because they cannot assign opacity in units smaller than 1 fully opaque pixel. For Dip 5, SART is clearly the best match to the shadow image; arc averaging produces a light curve which is too narrow, likely because the near-transparent pixels farther from the center which would have produced the ``wings" of the transit event have been averaged away in the last combination-and-normalization step of the algorithm. Meanwhile, for Dip 8, the shadow images from both arc averaging and SART match the light curve well.

We strongly caution that there is no straightforward way to interpret these images, for two reasons. First, these images are subject to both the flip and arc degeneracies; second, the grid resolution is low (N = 5 for Dip 5; N = 6 for Dip 8) because we are limited by the 30-minute cadence of \textit{Kepler} observations, and technically this resolution is so low that the small-planet limb-darkening approximation used to calculate the light curves of these shadow images is inappropriate. Nevertheless, these limitations should only affect the distribution of opacity among the semi-opaque pixels in the shadow images of Figures~\ref{fig:boyajian_dip5} and~\ref{fig:boyajian_dip8}; pixels which are fully transparent in the shadow images should remain so, even if we were to obtain a much higher-resolution time series of these events.

We therefore note that the ``gaps" of near-zero opacity (i.e., nearly transparent regions) which symmetrically frame the opaque transiting blob at the center of the shadow images in Figures~\ref{fig:boyajian_dip5} and~\ref{fig:boyajian_dip8} suggest that structured occulters are responsible for Dips 5 and 8 of the light curve of Boyajian's Star. The gap structure appears to be necessary to produce a shadow image which matches the ingress and egress shape of both Dip 5 and Dip 8; we note, for example, that this gap structure is missing in the arc-averaged shadow image of Dip 5 at $v = 3.2 \mathrm{d}^{-1}$, and the ingress and egress features of the corresponding light curve are too sharp to match the observed light curve.

\section{CONCLUSIONS}
\label{sec:conclusions}

Here, we have developed a mathematical and computational framework to address the problem of shadow imaging, or inferring the shape of a transiting object from its light curve alone. We find that this problem, which amounts to reconstructing a two-dimensional map from a one-dimensional time series, is degenerate, like the analogous problems of eclipse mapping and starspot inversion. In particular, by the flip degeneracy, shadow images are horizontally symmetrical; by the arc degeneracy, any infinitesimal opaque point in a shadow image can be replaced by a pair of intersecting semicircular arcs without consequence to the light curve; and by the stretch degeneracy, a wide image transiting at high velocity can produce the same light curve as a narrow image transiting slowly, given high enough pixel resolution.

In spite of these degeneracies, we are able to recover informative shadow images by adopting additional assumptions in algorithmic approaches to inverting light curves. We investigate four algorithms with different underlying assumptions. The first is arc averaging, which assumes that opacity should be distributed along arcs in inverse proportion to the $\sin{\theta}$ opacity distribution characteristic of the arc degeneracy. The second is the Simultaneous Algebraic Reconstruction Technique, an iterative approach which assumes that opacity should be distributed so as to minimize the RMS averaged over all time steps of the light curve and all pixels in the grid. The third is arc parsimony, which assumes that opacity should be distributed to as few individual opaque pixels as possible. The fourth is arc combinatorics, which assumes that opacity should be assigned to the best combination of individual opaque pixels to match the light curve. More broadly, the first two algorithms require only that the grid opacities be physical (i.e., restricted to the range $[0,1]$), while the latter two algorithms operate under the more restrictive assumption that the grid pixel opacities ought to be binary-valued. The less-restrictive case can accommodate pixel images of dusty, translucent, or solid objects smaller than the pixel scale, while the more-restrictive case is in principle more appropriate for recovering an image of a solid body which is larger than the pixel scale..

Overall, we conclude that SART is the approach which is most robust to our choices of grid resolution and velocity, most robust to noise in the observed light curve, and best able to accommodate shallow transit events. The only downside of SART is that, because it is an iterative optimization method, it is not parallelizable. For grids of the size investigated here ($N \leq 16$), it is of perfectly manageable computational cost.

We evaluate the performance of the four algorithms on a number of test cases, and find that we can recover informative shadow images for both binary- and continuous-valued opacity grids. We also apply them to real transit events---first, the triple transit of TRAPPIST-1 c, e, and f, for which the true shadow image is known. We recover a shadow image of TRAPPIST-1 c, e, and f which matches our expectations, subject to the constraint that our model grid transits the star at a single velocity, while the real TRAPPIST-1 planets all move individually. 

We also apply our techniques to two of the dips observed in Boyajian's Star, for which the true shadow image is unknown. We recover images which are self-consistent in the sense that the results from all four algorithms are qualitatively similar; also, the shadow images of Dip 5 and Dip 7 resemble each other. Transparent gaps in the shadow images of both events suggest that both dips were caused by structured occulters. However, we caution that these shadow images are difficult to interpret: they are subject to both the flip and arc degeneracies, and they are limited in resolution by the cadence of the original \textit{Kepler} observations. In the future, for successful shadow imaging of events like these, high time sampling of the light curve is key.

An important next step in shadow imaging will be to expand the framework presented here to encompass a true \textit{inference} of shadow images: in other words, to recover, given a transit light curve, a distribution of images which could have generated it, complete with uncertainties on the pixel opacities. Such a distribution would meaningfully represent the full set of degenerate solutions that could generate a particular observed set of uncertain flux measurements in a way that a single image cannot.

Accounting for measurement uncertainties is certainly possible within the work presented here; one could, for example, draw repeated ``realizations" of a particular light curve given the uncertainties on the individual flux measurements, then invert each realization to recover a single shadow image. The deeper question is how to take what is currently a deterministic retrieval procedure---one light curve, inverted with any of our algorithms, yields exactly one reproducible shadow image---and build in a way to account for the physical degeneracies of the problem, particularly the arc degeneracy, such that one light curve can generate an ensemble of possible shadow images.

In principle, one could also attempt to engineer such an ensemble from a single shadow image by perturbing opacity along arcs. We find that in practice, because of the complex overlapping pattern of the ingress and egress arcs, it is very difficult to perturb opacities and maintain a good fit to the observed light curve. In other words, the arc structure renders the pixel opacities strongly and non-trivially correlated. It remains nevertheless an interesting avenue for future work.

To accompany this work, we present the software package \texttt{EightBitTransit}, implemented in \texttt{Python}, which is able to calculate the light curves of arbitrary pixel arrangements and to recover shadow images from an input light curve, given the user's choice of grid parameters and inversion algorithm. This software package is available at \url{https://github.com/esandford/EightBitTransit}.

\acknowledgements
The  authors  thank the referee for a thorough and thoughtful review, and members  of  the  Cool  Worlds Lab  for  useful  discussions. ES thanks Zephyr Penoyre for help building the mathematical formalism of the arc degeneracy, and for many conversations throughout the project. ES thanks Moiya McTier for test-driving the \texttt{EightBitTransit} installation instructions.

\bibliography{mainbib}

\begin{thebibliography}{}
\expandafter\ifx\csname natexlab\endcsname\relax\def\natexlab#1{#1}\fi

\bibitem[{Andersen \& Kak(1984)}]{andersen84}
Andersen, A., \& Kak, A. 1984, Ultrasonic Imaging, 6, 81

\bibitem[{{Berdyugina} \& {Kuhn}(2017)}]{berdyugina17}
{Berdyugina}, S.~V., \& {Kuhn}, J.~R. 2017, ArXiv e-prints, arXiv:1711.00185

\bibitem[{{Bodman} \& {Quillen}(2016)}]{bodman16}
{Bodman}, E. H.~L., \& {Quillen}, A. 2016, \apj, 819, L34

\bibitem[{{Boyajian} {et~al.}(2016){Boyajian}, {LaCourse}, {Rappaport},
  {Fabrycky}, {Fischer}, {Gandolfi}, {Kennedy}, {Korhonen}, {Liu}, {Moor},
  {Olah}, {Vida}, {Wyatt}, {Best}, {Brewer}, {Ciesla}, {Cs{\'a}k}, {Deeg},
  {Dupuy}, {Handler}, {Heng}, {Howell}, {Ishikawa}, {Kov{\'a}cs}, {Kozakis},
  {Kriskovics}, {Lehtinen}, {Lintott}, {Lynn}, {Nespral}, {Nikbakhsh},
  {Schawinski}, {Schmitt}, {Smith}, {Szabo}, {Szabo}, {Viuho}, {Wang},
  {Weiksnar}, {Bosch}, {Connors}, {Goodman}, {Green}, {Hoekstra}, {Jebson},
  {Jek}, {Omohundro}, {Schwengeler}, \& {Szewczyk}}]{boyajian16}
{Boyajian}, T.~S., {LaCourse}, D.~M., {Rappaport}, S.~A., {et~al.} 2016,
  \mnras, 457, 3988

\bibitem[{{Claret} {et~al.}(2012){Claret}, {Hauschildt}, \& {Witte}}]{claret12}
{Claret}, A., {Hauschildt}, P.~H., \& {Witte}, S. 2012, \aap, 546, A14

\bibitem[{{de Wit} {et~al.}(2012){de Wit}, {Gillon}, {Demory}, \&
  {Seager}}]{deWit12}
{de Wit}, J., {Gillon}, M., {Demory}, B.~O., \& {Seager}, S. 2012, \aap, 548,
  A128

\bibitem[{{Deeg}(2009)}]{deeg09}
{Deeg}, H. 2009, in Transiting Planets, Vol. 253, 388--391

\bibitem[{{Farr} {et~al.}(2018){Farr}, {Farr}, {Cowan}, {Haggard}, \&
  {Robinson}}]{farr18}
{Farr}, B., {Farr}, W.~M., {Cowan}, N.~B., {Haggard}, H.~M., \& {Robinson}, T.
  2018, ArXiv e-prints, arXiv:1802.06805

\bibitem[{{Gillon} {et~al.}(2017){Gillon}, {Triaud}, {Demory}, {Jehin}, {Agol},
  {Deck}, {Lederer}, {de Wit}, {Burdanov}, {Ingalls}, {Bolmont}, {Leconte},
  {Raymond}, {Selsis}, {Turbet}, {Barkaoui}, {Burgasser}, {Burleigh}, {Carey},
  {Chaushev}, {Copperwheat}, {Delrez}, {Fernandes}, {Holdsworth}, {Kotze}, {Van
  Grootel}, {Almleaky}, {Benkhaldoun}, {Magain}, \& {Queloz}}]{gillon17}
{Gillon}, M., {Triaud}, A. H.~M.~J., {Demory}, B.-O., {et~al.} 2017, \nat, 542,
  456

\bibitem[{{Goncharskii} {et~al.}(1982){Goncharskii}, {Stepanov}, {Khokhlova},
  \& {Yagola}}]{goncharskij82}
{Goncharskii}, A.~V., {Stepanov}, V.~V., {Khokhlova}, V.~L., \& {Yagola}, A.~G.
  1982, \sovast, 26, 690

\bibitem[{{Juvan} {et~al.}(2018){Juvan}, {Lendl}, {Cubillos}, {Fossati},
  {Tregloan-Reed}, {Lammer}, {Guenther}, \& {Hanslmeier}}]{juvan18}
{Juvan}, I.~G., {Lendl}, M., {Cubillos}, P.~E., {et~al.} 2018, \aap, 610, A15

\bibitem[{{Katz}(2017)}]{katz17}
{Katz}, J.~I. 2017, \mnras, 471, 3680

\bibitem[{{Kawahara} \& {Fujii}(2011)}]{kawahara11}
{Kawahara}, H., \& {Fujii}, Y. 2011, \apj, 739, L62

\bibitem[{{Kipping}(2011)}]{kipping11}
{Kipping}, D.~M. 2011, \mnras, 416, 689

\bibitem[{{Kreidberg}(2015)}]{kreidberg15}
{Kreidberg}, L. 2015, \pasp, 127, 1161

\bibitem[{{Lanza} {et~al.}(1998){Lanza}, {Catalano}, {Cutispoto}, {Pagano}, \&
  {Rodono}}]{lanza98}
{Lanza}, A.~F., {Catalano}, S., {Cutispoto}, G., {Pagano}, I., \& {Rodono}, M.
  1998, \aap, 332, 541

\bibitem[{{Majeau} {et~al.}(2012){Majeau}, {Agol}, \& {Cowan}}]{majeau12}
{Majeau}, C., {Agol}, E., \& {Cowan}, N.~B. 2012, \apj, 747, L20

\bibitem[{{Mandel} \& {Agol}(2002)}]{mandelagol}
{Mandel}, K., \& {Agol}, E. 2002, \apjl, 580, L171

\bibitem[{{Metzger} {et~al.}(2017){Metzger}, {Shen}, \& {Stone}}]{metzger17}
{Metzger}, B.~D., {Shen}, K.~J., \& {Stone}, N. 2017, \mnras, 468, 4399

\bibitem[{Nelder \& Mead(1965)}]{neldermead65}
Nelder, J.~A., \& Mead, R. 1965, The Computer Journal, 7, 308

\bibitem[{{Piskunov} {et~al.}(1990){Piskunov}, {Tuominen}, \&
  {Vilhu}}]{piskunov90}
{Piskunov}, N.~E., {Tuominen}, I., \& {Vilhu}, O. 1990, \aap, 230, 363

\bibitem[{{Sing}(2010)}]{sing10}
{Sing}, D.~K. 2010, \aap, 510, A21

\bibitem[{{Vogt} \& {Penrod}(1983)}]{vogt83}
{Vogt}, S.~S., \& {Penrod}, G.~D. 1983, Publications of the Astronomical
  Society of the Pacific, 95, 565

\bibitem[{{Vogt} {et~al.}(1987){Vogt}, {Penrod}, \& {Hatzes}}]{vogt87}
{Vogt}, S.~S., {Penrod}, G.~D., \& {Hatzes}, A.~P. 1987, \apj, 321, 496

\bibitem[{{Wright} {et~al.}(2016){Wright}, {Cartier}, {Zhao}, {Jontof-Hutter},
  \& {Ford}}]{wright16megastructures}
{Wright}, J.~T., {Cartier}, K. M.~S., {Zhao}, M., {Jontof-Hutter}, D., \&
  {Ford}, E.~B. 2016, \apj, 816, 17

\bibitem[{{Wright} \& {Sigurdsson}(2016)}]{wright16}
{Wright}, J.~T., \& {Sigurdsson}, S. 2016, \apj, 829, L3

\end{thebibliography}

\listofchanges

\end{document}